\documentclass[useAMS,usenatbib]{mn2e}
\usepackage{graphicx}

\title[The hierarchical build-up of the Tully-Fisher relation]
      {The hierarchical build-up of the Tully-Fisher relation}

\author[C. Tonini et al.]
{Chiara Tonini$^{1,}$$^{2}$
\thanks{E-mail:ctonini@unimelb.edu.au},
Claudia Maraston$^{2}$,
Bodo Ziegler$^{3}$,
Asmus B\"ohm$^{4}$, 
\newauthor
Daniel Thomas$^{2}$,
Julien Devriendt$^{5}$
and Joseph Silk$^{5}$ \\
$^{1}$Centre for Astrophysics and Supercomputing, Swinburne University
of Technology, VIC 3122, Melbourne, Australia\\
$^{2}$Institute of Cosmology and Gravitation, University of Portsmouth, PO1 3FX Portsmouth, UK\\
$^{3}$ESO, Karl-Schwarzschild-Strasse 2, 85748 Garching bei Muenchen, Germany\\
$^{4}$Institute of Astro- and Particle Physics, Technikerstrasse 25/8, 6020 Innsbruck, Austria\\  
$^{5}$University of Oxford, OX1 3PU Oxford, UK\\
}

\begin{document}

%\date{Accepted 1988 December 15. Received 1988 December 14; in original form 1988 October 11}
  
%\pagerange{\pageref{firstpage}--\pageref{lastpage}} \pubyear{2009}

\maketitle

\begin{abstract}
We use the semi-analytic model GalICS to predict the Tully-Fisher relation
in the B, I and K bands, and its evolution with redshift, up to $z \sim 1$.
We refined the determination of the disk galaxies rotation velocity, with a dynamical
recipe for the rotation curve, rather than a simple conversion from the total mass
to maximum velocity. The new recipe takes into account the disk shape factor, and the angular momentum
transfer occurring during secular evolution leading to the formation of
bulges. This produces model rotation velocities that are lower by $\sim 30$ km/s
in case of Milky Way-like objects, and
$\leq 20-30$ km/s for the majority of the spirals, 
amounting to an average effect of $\sim 20-25 \%$.
We implemented stellar population models with a complete treatment 
of the thermally pulsing asymptotic giant branch,
which leads to a revision of the mass-to-light ratio in the near-IR. 
Due to this effect, K band luminosities increase by $\sim 0.5$ at redshift $z=0$
and by $> 1$ at $z=3$, while in the I band at the same redshifts 
the increase amounts to $\sim 0.3$ and $\sim 0.5$ mags.
With these two new recipes in place, the comparison between the predicted Tully-Fisher relation
with a series of datasets in the optical and near-infrared, 
at redshifts between 0 and 1, is used as a diagnostics of the 
assembly and evolution of spiral galaxies in the model.
At redshifts $0.4<z<1.2$ the match between the new model and
  data is remarkably good, expecially
for later-type spirals (Sb/Sc).
At $z=0$ the new model shows a net improvement in comparison with its
original version of 2003, and in accord with recent observations in
the K band, the model Tully-Fisher also
shows a morphological differentiation. However, 
in all bands the $z=0$ model Tully-Fisher is too bright.
We argue that this behaviour is caused by inadequate star formation histories in 
the model galaxies at low redshifts. The star-formation rate
declines too slowly, due to continuous gas infall that is not efficiently
suppressed. An analysis of the model disk scale lengths, at odds with
observations, hints to some missing physics in the modeling of disk
formation inside dark matter halos.
\end{abstract}


\section{Introduction}

The theory of galaxy formation has advanced remarkably in the past decade, 
thanks to the employment of sophisticated methods of investigation such as
numerical simulations and semi-analytic models (see for instance Balland et al. 2003, 
Baugh et al. 2005, Bower et al. 2006, 
Cattaneo et al. 2008, Cole et al. 2000, De Lucia et al. 2004, Firmani \& Avila-Reese 1999, Hatton et al. 2003, 
Kauffmann et al. 1993, Menci et al. 2006, Monaco et al. 2007, Somerville et al. 2008, 
Trujillo-Gomez et al. 2010, van den Bosch 2000, van den Bosch 2002). 
With these tools, the problem of the assembly of structures in the Universe can be addressed, 
and the vast number of non-linear physical processes that accompany the formation
and evolution of galaxies from primordial perturbations can be modelled. 

Interesting for galaxy formation are the observations of
scaling relations, that trace a pattern in the assembly history
of the vast variety of objects that we observe in the Universe. The origin
and nature of such relations is at the core of the mechanism of galaxy assembly
(Courteau et al. 2007).   
One of the most firmly established scaling relations is the Tully-Fisher 
(hereafter TF; Tully \& Fisher, 1977), that correlates the luminosity and 
the rotation velocity of disk galaxies.
The tightness and universality of the relation make it a valid tool to measure
extragalactic luminosity distances, since the rotational velocity does not depend on 
distance or cosmology. But a most interesting aspect of the relation is also the 
insight it provides on the mechanism of disk assembly (Tonini et al. 2006a). 

In the cold dark matter (CDM) scenario for disk galaxy formation (Fall \& Efstathiou, 1980),
the disk forms out of gas that cools inside a dark matter halo. Initially in equilibrium
with the dark matter, the gas collapses into a rotationally supported disk 
under angular momentum conservation,
while the halo evolves adiabatically (Blumenthal et al., 1986). As a consequence, the 
final equilibrium structure and dynamics of the disk are strongly correlated with those
of the dark matter halo, thus providing an insight into the physical halo parameters. 
In particular, the disk rotation velocity $V_{\rm D}$, the disk size parameterized through
the scale-length $R_{\rm D}$, and the disk mass $M_{\rm D}$ are determined by the halo virial 
properties, $V_{\rm vir} \propto R_{\rm vir} \propto M_{\rm vir}^{1/3}$ (Dutton et al., 2007).
The slope of the relation is a function of the radius at which the velocity is measured,
mirroring the dark matter mass distribution as a function of galactocentric
distance (Yegorova \& Salucci, 2007).

Interestingly, although both dynamics and baryonic physics enter the TF, the relation
is remarkably tight. In particular, galaxies follow the \textit{baryonic} TF, which links the
dynamical properties of disk (stars $+$ gas) and dark matter halo, with
a very small scatter (see for instance McGaugh et al. 2009). In the luminosity $vs$ velocity
TF relation, the measured luminosity is given by the stellar
emission, and is therefore a proxy for the stellar component.
The star formation rate (SFR) and the feedback processes,
that regulate the balance between stellar and gas mass, follow a scaling with the
dynamical quantities, and provide at the same time a source of scatter. 

In galaxy formation models, the supernova feedback efficiency is one of the main 
parameters driving the slope and zero-point of the TF relation. 
In fact, being the circular velocity
of the disk linked to the halo virial velocity, and therefore to its escape velocity, 
it is also closely related to the ability of the galaxy to blow a wind. The balance 
between feedback and star formation sets the amplitude of the TF relation, and its variation
across the mass range affects the TF slope (Croton et al. 2006, de Lucia et al. 2004). 

Semi-analytic models so far do not have a good grasp of the TF relation. In fact, 
although both Croton et al. (2006) and de Lucia et al. (2004) 
reproduce the local I-band TF relation (Giovannelli et al. 1997), this match should
not be considered a success because they do
not model rotation curves in a physically justified way. De
Lucia et al., assumes the total mass profiles are isothermal
(i.e., constant circular velocity), while Croton et al.  assume
the observed rotation velocity is equal to the maximum circular
velocity of the dark matter halo (in the absense of galaxy
formation). Both of these assumptions are expected to
under-predict the true rotation velocities.

On the other hand, semi-analytic models which actually model the galaxy rotation curves 
do not match the TF relation. Among these, GalICS (Hatton et al. 2003) includes
baryons, Cole et al. (2000), Benson et al. (2003) and Benson \& Bower (2010)
include baryons, NFW halos (Navarro, Frenk \& White (1997) and halo contraction.

In addition, it has been pointed out that semi-analytic models cannot simultaneously account for the
mass and angular momentum distribution of disk galaxies in a gas-dynamical context 
(Courteau et al., 2007, van den Bosch et al. 2002b, Governato et al. 2007).
In particular for the modeling of the TF relation, 
sources of systematic errors other than the feedback implementation 
include the structure of dark matter halos inferred from
simulations (the halo concentration and density profile for instance), and the modeling
of the dynamics of cooling and baryonic collapse inside halos (see for instance Tonini 
et al. 2006a, 2006b, Piontek \& Steinmetz 2009). 

Of interest in the present paper, 
the predicted TF relation is also affected by the modeling of the galaxy rotational velocity, 
and the stellar population models implemented. In this work we improve both these recipes
in the GalICS model, and re-address the comparison with the observed TF relation. We consider
a broad spectral range from the optical to the near-IR, and in particular we perform
the comparison in the K band. We also compare the model and observed 
evolution of the TF up to redshifts $z \sim 1$, both in the optical and 
in the near-IR.

In current semi-analytic models, the rotation velocity of disk galaxies is a crude estimate
based on the virial velocity of the host dark matter halos (for instance 
de Lucia et al. 2004, Croton et al. 2006), or the sum of a halo $+$ stellar component 
in spherical symmetry (Hatton et al. 2003). These approximations do not take into account
the actual mass profile of the disk, the effects of angular momentum exchange in the 
galaxy, and the dark matter halo expansion due to dynamical friction,  
and introduce systematic errors both in the zero-point and in the slope of the predicted TF
(see Dutton et al. 2007). In support to this point, Monaco et al. (2007) also showed that
the observed baryonic TF relation can be matched by hierarchical models if 
the dark matter halo concentration is lower than predicted by $\Lambda$CDM simulations
(to $c=4$ for halo masses around $10^{12} \ M_{\odot}$). 

In the present work, we implement for the first time in the semi-analytic model GalICS
(Hatton et al. 2003) a recipe for the galaxy rotation
curve that takes into account the shape of the disk and the angular momentum exchange between
disk and bulge and between the galaxy and the dark matter halo. We do so by modifying the
disk and halo parameters following the recipe proposed by Dutton et al. (2007). 

The TF relation predicted by galaxy formation models is also determined by the 
conversion between stellar mass and light, which 
is performed through the use of stellar population models. In the original version
of GalICS (Hatton et al. 2003) the observed TF relation was not reproduced.
In two recent papers (Tonini et al. 2009, 2010) we showed how the predictions of 
semi-analytic models are greatly affected by the choice of the input stellar populations.  
In particular we showed how the use of the Maraston (2005; hereafter M05) models allow 
GalICS to reproduce the colours and near-IR luminosities of $z \sim 2$ galaxies, in the 
mass and age ranges accessible to hierarchical clustering. In this paper we address the impact
of the M05 models on the predicted TF relation in the optical and near-IR, 
and its evolution with redshift. 

This Paper is organized as follows. In Section 2 we introduce the galaxy formation model,
and we describe the theoretical recipe to derive the galaxy rotation curves from the dynamical 
properties of the galaxies and their host dark matter halos, the stellar population models employed,
and the method for the selection of model spirals, including the definition of a proxy 
for the morphological differentiation of these objects. In Section 3 we describe the data
samples used to make the comparison between model and observations. In Section 4 we describe the 
results for the $z=0$ Tully-Fisher, and its dependence on morphology. In Section 5 we address
the evolution of the TF with redshift. Finally, in Section 6 we discuss our results and conclude.

\section{The model}

We produce the model galaxies through the hybrid semi-analytic model
GalICS (Hatton et al. 2003), and we defer the reader to its original
paper for details on the dark matter N-body simulation and the 
implementation of the baryonic physics. In brief, the model builds up the galaxies
hierarchically, and evolves the metallicity consistently with the 
cooling and star formation history (with the new implementation of the chemical evolution by Pipino et
al., 2009). Feedback recipes for supernovae-driven
winds and AGN activity are implemented in the code (the latest with the improved
version of Cattaneo et al. 2008). Merger-driven
morphology evolution and satellite stripping and disruption are taken into account.
 
In order to compare the model output with the data, we produce mock galaxy catalogues 
both in rest-frame and observer-frame, adopting each time the broadband filter sets of
the data available for comparison. The broadband magnitudes thus obtained are further 
scattered with gaussian errors comparable to the 
observational errors of our data samples (on average $\sigma = 0.1$ mag).
Dust extinction is treated in the model, however the observed TF relations that 
we use for the comparison with data are corrected for internal extinction by the 
various authors. For this reason, we produce unreddened galaxies for the comparison
with rest-frame TF relations.
When comparing our model galaxies with observed-frame data (not corrected for reddening), 
we implement dust extinction with a Calzetti extinction curve
and a colour-excess $E(B-V)$
proportional to the star formation rate for each single galaxy, 
parameterized as $E(B-V) = 1/3 \cdot (\log (SFR)-2) + 1/3$ (see Tonini et al. 2010 for
a discussion about this choice). 
Magnitudes are presented in the AB system.
We calculate the galaxy rotational velocity at the standard value of $r=2.2R_{\rm D}$
(where $R_{\rm D}$ is the exponential disk scale-length), 
consistently with the observations used in our comparisons. 

The semi-analytic model is rendered for the parameter set $H_0=71.9 \ \rm km \ s^{-1} \ Mpc^{-1}$, 
$\Omega_{\Lambda}=0.742$ and $\Omega_{\rm M}=0.258$; these parameters are used to compute
observed-frame magnitudes.
The model absolute magnitudes are not directly dependent on the cosmology
(which obviously enters in the physics of the semi-analytic model, but not in the computation
of the rest-frame light emission from galaxies); when comparing with multiple
sets of data, analyzed by different authors with
various cosmologies, we re-scale all units to the model $h=H_0/100$. 

\subsection{Galaxy rotational velocity}

We calculate the rotational velocity of each galaxy by building its rotation curve
and evaluating it at $r=2.2R_{\rm D}$, where $R_{\rm D}$ is the disk scale-length. 
Galaxies in the models are systems 
composed of a dark matter halo and stars distributed in a disk and a 
bulge; neglecting the gas component (which is not dominant in mass, at least 
at $z<1$) the rotation curve is
\begin{equation}
V_{\rm rot}^2(r)=V^2_{\rm H}(r)+V^2_{\rm disk}(r)+V^2_{\rm bulge}(r)~, 
\label{v2}
\end{equation} 
with $r$ being the galactocentric distance.
For the spheroidal components, i.e. the halo and the bulge, the contribution
to the velocity is simply $V^2(r)=\sqrt{GM(r)/r}$. Following the original GalICS recipe,
the halo density profile is defined as a truncated isothermal sphere. In the simulation
halos are tidally truncated by the surrounding density field, and the inner region
of halos is virialized. For the purpose of building the rotation curves, we consider halos
as spherically symmetric isothermal spheres out to the virial radius. In addition, halos
have a finite density core to avoid the singularity. The density profile can be parameterized as follows:
\begin{equation}
\rho(r)=\frac{\rho_0}{\left( \frac{r}{r_{\rm S}} \right)^{\gamma}\left[ 1+\left(\frac{r}{r_{\rm S}} \right)^{1/\alpha} \right]^{\alpha (\beta - \gamma)}}~,
\label{sphere}
\end{equation}
with $(\alpha, \beta, \gamma)=(1,2,0)$. Fits to the radial mass distrubution of halos
in dark matter-only N-body simulations produce the NFW density profile (NFW 1997), which belongs to the
same family of functions, with parameters $(1,3,1)$. At intermediate radii NFW halos and isothermal
spheres have the same density profile, while at the centre the NFW is shallower and at the outskirts
it is steeper. However, in the range of galactocentric distances of interest in the present work, the difference
between the two profiles is negligible (as is illustrated later on in
this Section). In general, the choice of the halo functional form is a source of
uncertainty, and it should be noted that there is an ongoing debate in
the literature regarding the shape of dark matter halos in the
presence of baryons, expecially in the center of galaxies where the
dark matter component reacts to baryonic mass assembly with either
expansion or contraction (see for instance Tonini et al. 2006b, Duffy
et al. 2010).

The halo is almost entirely pressure-supported, with a mild spin component, parameterized by 
$\lambda=J |E|^{1/2}/G M_H^{5/2}$ in terms of the halo total mass $M$, angular momentum $J$ and 
binding energy $E$. 
In simulations the distribution of the halo spin parameter appears to be log-normal, with both median and
scatter independent of mass and redshift to a good approximation (see e.g. Bullock et al. 2001,
Vitvitska et al. 2002, Hetznecker \& Burkert 2006, Bett et al. 2007, Maccio et al. 2007).
The median spin parameter for GalICS relaxed halos is $\lambda \sim 0.04$.
For each galaxy in GalICS, all halo parameters (virial mass and radius, and velocity dispersions) 
are directly taken from the N-body simulation, and naturally cause some scatter in the halo profile.

The bulge stellar mass distribution follows a Hernquist (1990) profile, characterized 
by parameters $(1,4,1)$, and is assumed to be entirely pressure-supported (see Hatton et al. 2003 for details
on bulge formation and geometry).
 
The disk structure and rotation is determined by the halo spin parameter $\lambda$
following Mo et al. (1998). The disk forms from a baryonic gaseous component that
initially has the same specific angular momentum as the halo, a mass $m_{\rm D}$ in units
of the halo mass, and a total angular momentum $j_{\rm D}$ in units of the halo angular 
momentum. The disk is assumed to have an exponential
surface density profile, with a scale-length defined as: 
\begin{equation}
R_{\rm D} = \frac{1}{\sqrt{2}} \left( \frac{j_{\rm D}}{m_{\rm D}} \right) \lambda R_{\rm H}~, 
\label{rd}
\end{equation}
where $R_{\rm H}$ is the halo virial radius (see Mo et al. 1998 for details).
In galaxy formation models the assumption of  
specific angular momentum conservation during disk formation is widely used, 
which leads to $j_{\rm D}/m_{\rm D}=1$. This assumption, while useful to semplificate the problem
and avoid additional degrees of freedom, is often unrealistic. 
In fact, angular momentum transfer from the 
infalling baryons to the halo pushes the ratio lower than unity, and both the halo and the 
disk structure are be affected (see Tonini et al. 2006b; also, Piontek \& Steinmetz 2009).
On the other hand, the ratio can also become greater than unity in case of 
removal of low angular momentum material (e.g., Maller \& Dekel 2002, Dutton \& van den Bosch 2009).
A discussion on the disk sizes of the model galaxies is to be found in the last Section.

In the Hatton et al. (2003) version of GalICS, the disk component of the velocity
was treated as $V^2(r)=\sqrt{GM(r)/r}$ (for future reference, this will be called 
the \textit{'spheroidal'} velocity model).
However, neglecting the shape factor in the disk 
gravitational potential leads to an underestimation of the maximum rotational velocity
by about $15 \%$, under angular momentum conservation (Binney \& Tremaine, 2008).
For the disk component, in order to take into account the shape factor we assume that
\begin{equation}
V^2_{\rm disk}(r)=\frac{G M_{\rm D}}{2R_{\rm D}} \ \left( \frac{r}{R_{\rm D}} \right)^2 [I_{\rm 0}K_{\rm 0}-I_{\rm 1}K_{\rm 1}]_{r/2R_{\rm D}}~,
\label{vdisk}
\end{equation}
where $M_{\rm D}$ is the total disk mass, and $I,K$ are modified Bessel functions of first and
second kind (Tonini et al. 2006, Salucci et al. 2007). 
This recipe effectively increases the rotational velocity of disk-dominated systems,
and has therefore an effect on the TF relation expecially at the high-mass end. 
For future reference, we call
$V_{\rm disky}$ the total rotational velocity when $V^2_{\rm disk}$ is modeled as in 
Eq.~(\ref{vdisk}).

Angular momentum conservation is not a realistic scenario also during bulge formation, 
as pointed out by Dutton et al. (2007). 
When bulges form through disk instabilities, part of the material in the disk loses its 
angular momentum, leading to the formation of a pressure-supported component at the 
center. The lost angular momentum is transferred to the rest of the disk and the 
dark matter halo, as seen in numerical simulations (Hohl 1971, Debattista et al. 2006). 
This affects the value of $R_{\rm D}$ through $j_{\rm D}/m_{\rm D}$ after bulge formation. 
Following Dutton et al. (2007), after defining $\Theta=M_{\rm bulge}/M_{\rm disk}$, 
the new disk scale-length is 
\begin{equation}
R_{\rm D} = \frac{1}{\sqrt{2}} f_{\rm R} \ f_{\rm H} [1+(1-f_{\rm x})\Theta]\left( \frac{j_{\rm D}}{m_{\rm D}} \right) \lambda R_{\rm H}~;
\label{rdutton}
\end{equation} 
the quantity $f_{\rm x}$ expresses the ratio between the specific angular momentum lost 
by the disk to the halo during bulge formation, to the total specific angular momentum
of the baryonic material that went to form disk and bulge. 
We re-calculate the disk scale-lengths of the galaxies in the simulation with this new recipe, 
obtaining 
\begin{equation}
R_{\rm D_{\rm new}}=R_{\rm D_{\rm MMW}} \left( 1+(1-f_{\rm x}) \frac{M_{\rm bulge}}{M_{\rm disk}} \right)~,
\label{rdafter}
\end{equation}
and we use the fiducial value $f_{\rm x}=0.25$ indicated by Dutton et al. (2007).
By inserting this new value of $R_{\rm D}$ into Eq.~(\ref{vdisk}), we obtain what we will
refer to as the \textit{'new'} model for the total rotational velocity, $V_{\rm new}$, from
Eq.~(\ref{v2}).

\begin{figure}
\includegraphics[scale=0.4]{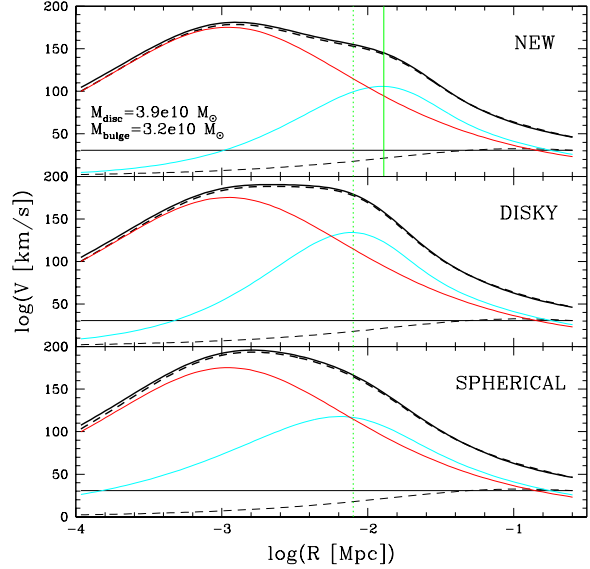}
\caption{Rotation curve of a simulated galaxy at $z=0$, with disk mass 
$M_{\rm disk}=3.9 \ 10^{10} \ M_{\odot}$ and bulge mass $M_{\rm bulge}=3.2 \ 10^10 \ M_{\odot}$.
The 'new', 'disky' and 'spheroidal' models for the rotation curve are compared (\textit{from upper to lower
panel}). The \textit{vertical dotted green line} shows the radius $r=2.2 R_{\rm D}$ before 
including the angular momentum transfer from bulge to disk. In the upper panel, the 
\textit{vertical solid green line} shows the new location of $r=2.2 R_{\rm D}$ after angular momentum transfer.
In all panels, \textit{solid thick black lines} show the total 
rotation curve,
\textit{solid blue lines} show the disk component of the velocity, \textit{solid red lines} show the 
bulge component. The \textit{solid thin black lines} show the halo component as a truncated isothermal sphere, 
while the \textit{dashed thin black lines} show the halo contribution as an NFW profile. The \textit{dashed 
thick black lines} show the total rotation curve if the halo is an NFW.}. 
\label{RC}
\end{figure}

Fig.~(\ref{RC}) shows the comparison between the three models for the rotation curve, 'new', 'disky' and
'spheroidal' (\textit{panels from upper to lower, respectively}), 
for a simulated galaxy at $z=0$, which was chosen for the conspicuous bulge 
($M_{\rm bulge}=3.2 \ 10^10 \ M_{\odot}$ and $M_{\rm disk}=3.9 \ 10^{10} \ M_{\odot}$). 
In all panels, the \textit{vertical dotted green line} shows the radius $r=2.2 R_{\rm D}$ before 
the evaluation of the angular momentum transfer from bulge to disk. In the upper panel, the new value of
$r=2.2 R_{\rm D}$ following the angular momentum transfer ('new' model) is shown as a 
\textit{vertical solid green line}. In all panels, \textit{solid thick black lines} show the total 
rotation curve,
\textit{solid blue lines} show the disk component of the velocity, \textit{solid red lines} show the 
bulge component. 

Note that, for a pure disk, the peak of the rotation curve is always approximately at $r=2.2 R_{\rm D}$
(making this particular radius the favourite in observations). 
From the 'spheroidal' to the 'disky' model, the disk component of the velocity features a more
distinguished peak, with the result that the maximum of the total rotation curve is closer to
$r=2.2 R_{\rm D}$, and the bulge is less prominent in determining its value. The sharpening of the
disk peak causes the measured total velocity to increase. 

However, the presence of other mass components, 
such as the bulge and the halo, tend to broaden the total velocity peak and push it to 
smaller radii, where these components are more concentrated. But the concentration of the bulge mass
at the expense of the disk must be accompanied by angular momentum transfer from the bulge to the disk itself, 
and according to Eq.~(\ref{rdafter}), the higher the ratio $M_{\rm bulge}/M_{\rm disk}$, 
the more the disk scale-lenght consequently increases, leading to the 'new' model and a 
re-distribution of the disk mass. The disk expands radially and, since it conserves its mass, 
it becomes less dense. Its velocity peak broadens and it is pushed outwards, thus
stretching the total rotation curve. Due to the decreased density, the disk potential well is
lower, and this, together with the rapidly declining bulge contribution, lowers the total
velocity measured at the new $r=2.2 R_{\rm D}$.
Since this effect if brought about by the presence of the bulge, it is going to be expecially
important at the high-mass end of the model TF relation, where bulge formation is more frequent
(this is a feature of the hierarchical model which will be discussed later on).

In Fig.~(\ref{RC}) we also show the halo component of the total rotation curve. 
The \textit{thin solid black line} is the velocity profile of the truncated isothermal sphere,
while the \textit{thin dashed black line} is the velocity profile of the NFW halo of same 
virial parameters. The total rotation curve obtained with the NFW (\textit{thick solid black lines})
is virtually undistinguishable from the one obtained with the truncated isothermal sphere 
(\textit{solid thick black lines}), expecially around the region $r=2.2 R_{\rm D}$ and beyond.

The new recipe for the rotation velocity allows for a sharper differentiation of the model
rotation curves depending on the galaxy morphology, a fact that has been long since observed
in nature (Salucci et al. 2007). In particular, in the upper panel of Fig.~(\ref{RC}) 
the bump in the rotation curve due to the presence of the bulge component is clearly visible.

The velocity recipe can be further optimized if one takes into account the dark matter 
halo evolution triggered by the formation of the galaxy. As proposed in Tonini et al. (2006b)
and discussed in Dutton et al. (2007), angular momentum exchange between the collapsing
baryonic component and the dark matter causes halo expansion, which affects the halo 
structural shape and ultimately the dark matter contribution to the rotation curve.
However, such a recipe would require more fundamental modifications
to the prescriptions for the dark matter halos in the semi-analytic code, and is
beyond the scope of this paper.

\begin{figure}
\includegraphics[scale=0.4]{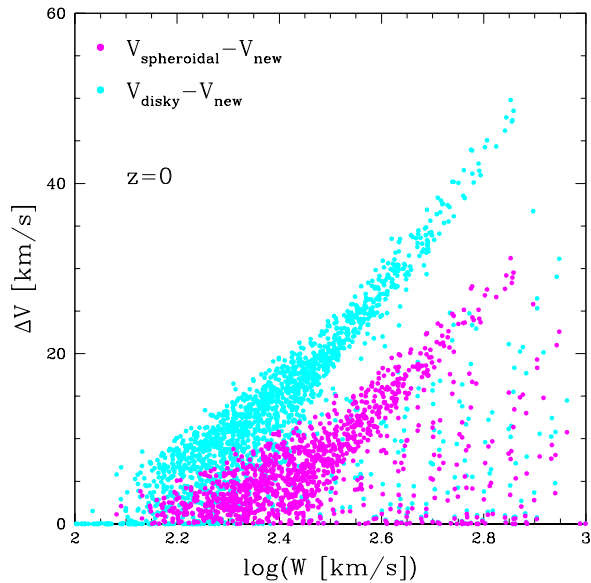}
\caption{The difference in the predicted $z=0$ galaxy rotational velocity at $r=2.2R_{\rm D}$ with different
assumptions for the disk velocity, as a function of the total 'new' rotation velocity. 
\textit{Magenta points}: difference between 
the original Hatton et al. (2003) model and the 'new' model.
\textit{Cyan points}): difference between the 'disky' model and the 'new' model presented here.
For each of the models, we plot the total velocity difference $\Delta V$ 
against the quantity $log(W=2V_{\rm rot})$,
which is used for the comparison with observations later on.}
\label{diffv}
\end{figure}

Fig.~(\ref{diffv}) illustrates the difference in the predicted total rotational velocity (at $z=0$)
at the radius $r=2.2R_{\rm D}$ between the 'new' velocity model and the 
'spheroidal' model (\textit{magenta dots}), 
and between the 'new' model and the 'disky' model (\textit{cyan dots}), as a function
of $log(W=2V_{\rm new})$, which is used for the comparison with observations later on. 

The combined effect of increasing the disk scale-lenght and decreasing the bulge contribution
to the rotation curve, is that the 'new' model velocity is always lower than the 'spheroidal'
model velocity, with a difference up to $\sim 30$ km/s at the high-mass end. On the other hand,
the 'new' model velocity is also always lower than the 'disky' model velocity, by up to $\sim 50$ km/s
at the high-mass end, where bulges contribute significantly to the stellar mass. The difference $\Delta V$
correlates in fact strongly with the total mass, and  
tends to zero for galaxies with small bulges.

With the new recipe in place, model rotation velocities are 
lower by $\sim 20-30$ km/s or less, compared to the 'spheroidal' model, for 
spirals less massive than the Milky Way, and by $\sim 40-50$ km/s for Milky Way-like objects.

In what follows, we adopt the 'new' model to compare the model predictions with
the data.

\subsection{Stellar population models}

\begin{figure}
\includegraphics[scale=0.4]{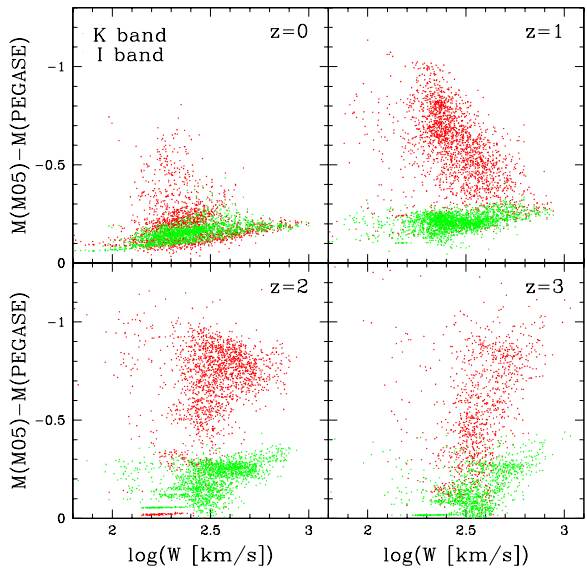}
\caption{Difference between the predicted rest-frame TF in the M05 and PEGASE runs, 
as a function of redshift. The \textit{green dots} show the I band and
the \textit{red dots} show the K band.}
\label{diff}
\end{figure}

The photometry of the model galaxies is produced with the use of the M05
stellar population models, which include a detailed treatment of the Thermally Pulsing
Asymptotic Giant Branch phase of stellar evolution. The implementation of the M05 
stellar populations in the semi-analytic model significantly improves the predictions 
for the colours and near-IR luminosities of galaxies, expecially at high redshift. 
The comparison of the model performance when equipped with the M05 and with its
original input stellar populations (PEGASE, Fioc \& Rocca-Volmerange, 1997) shows
that the inclusion of the TP-AGB produces near-IR luminosities more than 1 magnitude
brighter for a given stellar mass, and $V-K$ colours more than 1 magnitude redder,
matching the observations of high-redshift galaxies 
(Tonini et al. 2009, 2010, Henriques et al. in prep.). 

The TP-AGB light emission starts at wavelengths $\lambda > 5000 \ \mathring{A}$,
peaks in the rest-frame K band, and affects the nearby bands
from the red part of the optical spectrum down to the infrared. This affects both the
zero-point and the slope of the predicted TF relation, as a function of redshift. 
Fig.~(\ref{diff}) shows the difference in magnitude between the TF relation predicted with the M05 and
the PEGASE runs of the semi-analytic model, in the rest-frame I band (\textit{green dots})
and K band (\textit{red dots}), in four redshift bins from 0 to 3. On the $x-$axis we plot
$W=2 V_{\rm rot}$.
The difference between the M05 and PEGASE TF in the I band tends to mildly increase with
increasing redshift. The median magnitude
difference in the redshift bins centered in $z=0,1,2,3$ is respectively 
$\Delta(TF)\sim 0.15, 0.22, 0.25, 0.2$ mags. At $z=2,3$ there is a mild dependence with
galaxy mass, so that the difference can go up to $\Delta(TF)\sim 0.35-0.4$ for massive galaxies.
In the K band, we find a significant increase of the difference $\Delta(TF)$ with redshift. 
In the redshift bins centered in $z=0,1,2,3$ we find the median $\Delta(TF)\sim 0.2, 0.65, 0.8, 0.7$ mags
with maximum $\Delta(TF)> 1$ mags for $z \ge 1$. There is no significant trend with galaxy mass.

From now on in this work, the M05 models will be used for all the comparison with data.

\subsection{Selection of model galaxies}

The determination of the morphology of the model galaxies is based
on the relative contribution of bulge and disk to some galactic property, 
and can be carried about in various ways.
For the purpose of a more comprehensive study, we adopted two different
criteria for the selection of the model spiral galaxies.

Hatton et al. (2003) defined as spiral galaxies the objects for which the ratio 
between the bulge and disk luminosity in the B band did not exceed a 
given threshold. Simien \& de Vaucouleurs (1986) found that this
ratio correlates well with the Hubble type (see for instance Graham
\& Worley 2008 for an updated study). Following this criterium,
if we define the parameter $\gamma = exp(-L_{\rm bulge}/L_{\rm disk})$, the objects with
$\gamma > 0.507$ are defined as spirals (see also Baugh et al. 1996).  
This classification has the advantage to be observationally-oriented, but it depends
on the spectrophotometric model in use (including dust reddening), 
and yields no dynamical information.

From the semi-analytic model point of view, we have the information about the mass and size
of disk and bulge \textit{before} we compute the stellar emission, so we can 
also classify our galaxies directly in terms of the ratio between the bulge and 
disk \textit{stellar mass}, to mimic the morphological differentiation. 
We choose to split our galaxies in 
5 subsamples, defined by $M_{\rm bulge}/M_{\rm disk}$ in 5 intervals:
galaxies with $M_{\rm bulge}/M_{\rm disk}<0.25$ loosely correspond to Sc/Irregulars, 
galaxies with $0.25<M_{\rm bulge}/M_{\rm disk}<0.5$ loosely correspond to Sb spirals, galaxies with
$0.5<M_{\rm bulge}/M_{\rm disk}<0.75$ loosely correspond to Sa spirals, and galaxies with 
$0.75<M_{\rm bulge}/M_{\rm disk}<1$ are similar to 'S0' types. Strictly speaking, 
observed S0 galaxies are not identified depending on the bulge/disk mass ratio, but
are rather selected to be passively evolving disks, based on colours.
Our morphology selection does not depend on photometry, but nonetheless it naturally
preserves the correspondence between bulges and old stellar populations, since 
bulges in the model do not form new stars (except in merger events).

We find that our somewhat arbitrary selection criterium qualitatively matches the
observed properties of spirals (K. Masters, private communication). 
It is also consistent with 
other prescriptions used in the literature: we re-selected our Sb/Sc spirals
following both Croton et al. (2006) and de Lucia et al. (2004), and found
that the resulting subsample coincides with the one selected in terms of 
$M_{\rm bulge}/M_{\rm disk}$. The totality of the spirals in the range $0 < M_{\rm bulge}/M_{\rm disk} < 1$
is almost coincident with the sample selected with the Hatton et al. (2003)
criterium (with the exception of some massive bulge-dominated objects that Hatton selected
as spirals). As $M_{\rm bulge}/M_{\rm disk}$ increases, the average age of the stellar populations 
tends naturally to increase.

The hierarchical nature of the semi-analytic model causes a large number of 
spiral galaxies to be subject to some kind of perturbations (like gas stripping
and strangulation) when they live in dense regions, with the consequent quenching
of star formation. In any galaxy identified as a satellite, i.e. infalling into the halo
of a central galaxy (which on the contrary sits in the center of its halo), cooling
of fresh gas is shut down, unless received via a merger. A comparison of the model satellites
with observed spirals is therefore improper, and as will be shown in Section 4, largely
unsuccessful. For this reason, we use only model central galaxies to build the TF.

\section{Data selection}

The data at our disposal are either in rest-frame or observed-frame.
When data are in rest-frame, the absolute magnitudes, evolutionary and $k$-corrections, and 
corrections for internal extinction were performed by the various authors. 
We compare these sets of data with unreddened model galaxies.
When data are in observed-frame and uncorrected for dust extinction, we compare
them with model galaxies to which we apply our reddening recipe.

\subsection{Local data}

For the $z=0$ Tully-Fisher relation in the I and K bands, 
we use the data by Masters et al. (2006, 2008).
Their I-band sample, called Spiral Field I++ is a compilation of objects taken 
from the Spiral Field I band and 
the Spiral Cluster I band catalogues analyzed by 
Giovanelli et al. (1994, 1995, 1997a,b) and Haynes et al. (1999a,b),
the SC2 sample (Dale 1998, Dale et al. 1999), data from Vogt (1995)
and Catinella (2005) and the HI archive from Springbob et al. (2005).
The SFI++ contains $~5000$ galaxies, and Masters et al. (2006) analyze 
a subsample of data consisting of spirals in the vicinity of nearby clusters.
The rotational widths are derived from HI global profiles when possible (about $60\%$ of the
sample), or optical rotation curves. See Masters et al. (2006) for 
details about the velocity derivation.
These objects are specifically targeted for I-band observations, and although
the sample is not complete (in magnitude or volume), it 
includes all types of spirals. However, as a result of various bias corrections, 
the sample is designed to be dominated
by late-type spirals (see Masters et al. 2006). The derived TF relation
is consistent with the one published by Giovannelli et al. (1997). 

For the K band, the data come from the 2MASS Tully-Fisher Survey (2MTF), 
cross-matched with the 2MASS Extended Source Catalog (XSC) including
all the galaxies with rotational widths coming either from the Cornell HI 
digital archive (Springbob et al. 2005) or the Cornell 
database of Optical Rotation Curves (ORCs) described by Catinella (2005).
The cross-match was performed for objects in the vicinity of the clusters 
in the SFI++ sample. 
About $65\%$ of the galaxies in the 2MTF sample belong also to the SFI++ sample.
Details about the corrections applied to the rotational widths, to account for
inclination, turbulent motions, cosmological broadening and instrumental effects
can be found in Masters et al. (2008), along with a detailed list of 
all the bias corrections. 

For both the I and K band samples described above, Masters et al. (2006,2008) find a 
morphological dependence of the slope and zero-point of the Tully-Fisher relation.

Further data for the I band, together with the B band at $z=0$, come from Verheijen (2001).
In this case the sample belongs to the Ursa Major cluster (Tully et al. 1996). Only galaxies
with a lower inclination limit of $45^{\circ}$ are considered, which leaves a sample of
49 galaxies described as the \textit{complete} sample. Corrections
for internal extinction are treated following Tully et al. (1998).
In the complete sample, the rotation velocity is estimated from the width of HI lines, 
corrected for instrumental resolution, internal turbulent motions and inclination. 
Further subsamples are defined according to the observed properties of the galaxies; one of these
subsamples is called the Rotation Curve sample (RC), containing galaxies 
for which synthesis imaging data is available, and the shape of the HI rotation curve is measured.
In this case, velocity is also estimated at the maximum of the rotation curve ($V_{\rm max}$)
or in the flat part of the curve ($V_{\rm flat}$) depending on the curve shape; these 
estimates are in good agreement with the HI-width method (see Verheijen 2001 for details). 
We also use the relation derived in the B band by Pierce \& Tully (1992) for comparison. 

\subsection{High-redshift data}

The evolution of the B-band TF relation up to $z \sim 1.3$ is compared with data 
from Fernandez-Lorenzo et al. (2009), 
in a sample belonging to the DEEP2 survey (Davis et al., 2003). The photometric catalogue (Data Release 1,
Coil et al. 2004) features magnitudes corrected for dust extinction (after Tully et al. 1998), 
and is $k$-corrected following Blanton \& Roweis (2007).
The rotation velocity has been estimated from optical line widths of integrated spectra, obtained
through line-fitting techniques. The selection of spiral galaxies was performed 
through visual inspection of HST images of the AEGIS survey, where it overlaps with the
DEEP2. Inclination angles of the selected sample are all between $25^{\circ}$ and $80^{\circ}$.
The sample was purged of E/S0 galaxies and interactive pairs, and includes spirals and irregulars. 
For comparison, we also take the published TF of Bamford et al. (2006) for field galaxies
at $<z>=0.33$.

The K-band TF up to $z \sim 1.2$ is compared with data from Fernandez-Lorenzo et al. (2010).
The sample belongs to the Groth Strip Survey, with spectroscopy provided by the DEEP2 survey 
(Data Release 3). The photometric data are part the AEGIS survey (Davis et al. 2003, 2007); the K-band
data were collected with Palomar WIRC (Bundy et al. 2006), and with the Infrared Array Camera on 
the Spitzer Space Telescope (Barmby et al. 2008). Data are corrected for internal extinction, and
galaxy morphology is selected by eye on Hubble Space Telescope images. As in 
Fernandez-Lorenzo et al. (2009), the rotational velocity is determined through optical line widths.

Another sample of optical and near-IR data comes from B\"ohm \& Ziegler (2007). 
It contains 73 galaxies with photometry in 7 bands from the UV to the near-IR.
This observational sample was constructed utilizing the multi-band imaging survey
of the FORS Deep Field (FDF, Heidt et al. 2003), which comprises
deep $U$, $B$, $g$, $R$, $I$, $J$ and $K$ photometry performed with the
Very Large Telescope and the New Technology Telescope.
To extract rotation curves from the 2D spectra, the $\lambda$-positions
(linecenters) of a given emission line were measured by fitting
Gaussians stepwise along its spatial profile.
Due to the small apparent sizes of the galaxies, the observed
rotation curves are heavily blurred.  
At redshifts of $z \approx 0.5$, the apparent scale lengths are of the
same order as the slit width and the seeing disk.
To correct for these blurring effects, 
synthetic rotation velocity fields were generated assuming an intrinsic shape
with a linear rise of the rotation speed at small radii and a turnover
into a regime of constant rotation velocity $V_{\rm max}$ at large radii
(for details on this prescription as well as tests of various rotation curve 
shapes, see B\"ohm et al. 2004).
Intrinsic dust absorption was corrected following Tully \& Fouqu\'e (1985).

\section{z=0 Tully-Fisher relation}

In this Section we present the $z=0$ model TF compared to the observational
relations in the B, I and K photometric bands. This choice of bands
is functional to test the overall performance of the model in relation to the 
star formation rates and the balance between stellar populations of different
ages in our galaxies, as well as the contribution of the TP-AGB light in 
the near-IR. 
We show the morphological differentiation of the K-band TF relation, which is predicted in
the model and present in the data. Moreover, we analyze the relative
contributions of our dynamical recipe and the stellar population models in 
determining the K-band TF relation. We compare the performance of
the model for central galaxies and satellites. We also discuss a possible origin for
S0 galaxies. 
 
\subsection{Tully-Fisher relation in B, I and K bands}

\begin{figure*}
\includegraphics[scale=0.9]{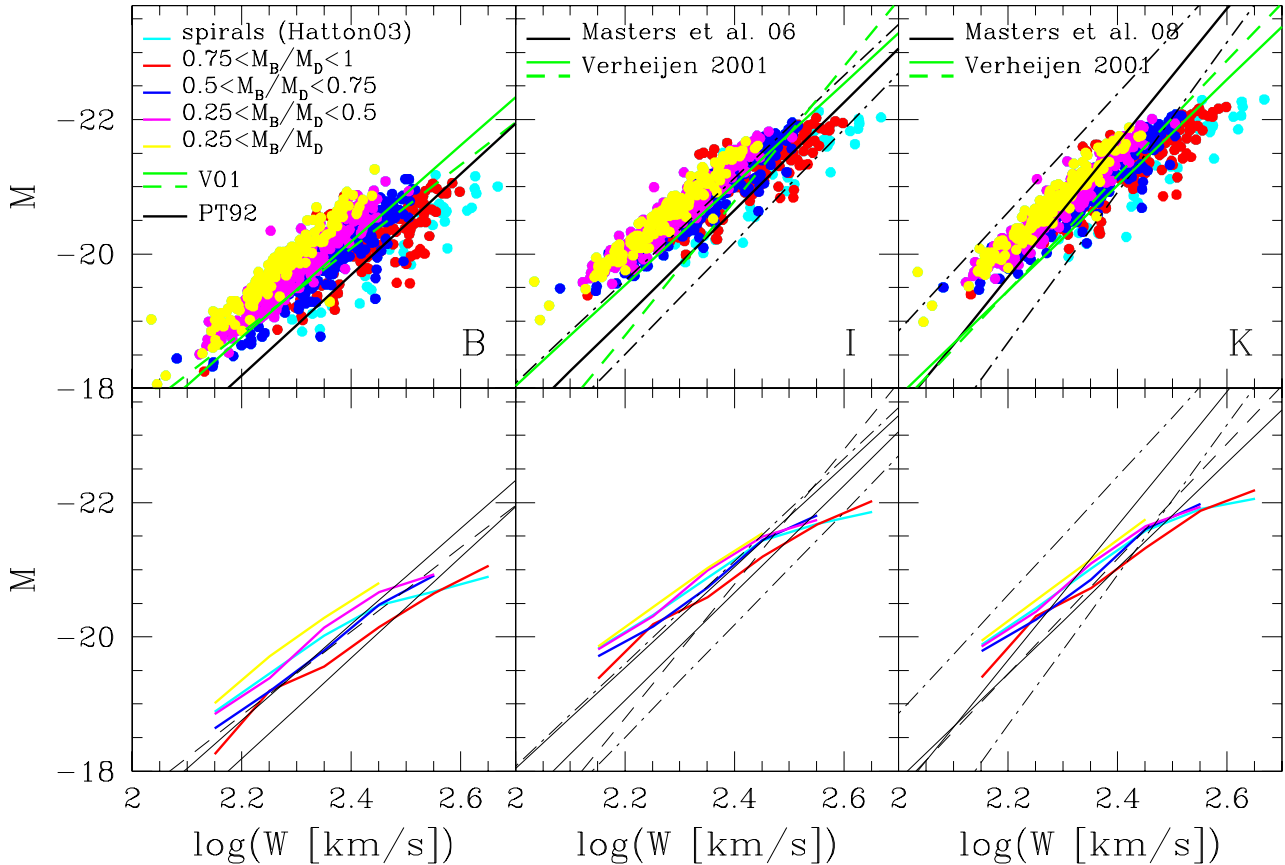}
\caption{The $z=0$ Tully-Fisher relation in the B, I and K bands (\textit{from left to right}). 
The \textit{upper panels} show the scatter plots, while the \textit{lower panels} show the median 
relations. 
In all panels, model spiral galaxies are selected according to $M_{\rm bulge}/M_{\rm disk}$ (see text).
(\textit{Left panel}): the B-band model TF (\textit{left panel}) is compared with the 
observed relation by Verheijen (2001,
\textit{green solid/dashed lines}, see text) and Pierce \& Tully (1992; 
\textit{solid black line}). (\textit{Central panel}): the I-band model TF is compared with 
the observed one from Masters et al. 2006 (\textit{black solid line} for the
relation, and \textit{dot-dashed lines} for its scatter) and Verheijen 2001
(\textit{green solid/dashed lines}).
(\textit{Right panel}): the K-band model TF relation is compared with 
the observed relation by Masters et al. (2008; \textit{solid black line} for the
relation, and \textit{dot-dashed lines} for its scatter) and Verheijen 2001
(\textit{green solid/dashed lines}).} 
\label{tfz0}
\end{figure*}

Figure (\ref{tfz0}) shows the predicted $z=0$ Tully-Fisher relation in the B, I and K bands
(\textit{from left to right}); scatter plots are shown in the \textit{upper panels}, while the 
median relations are shown in the \textit{lower panels}. 
Disk galaxies in the model are selected according to the 
$M_{\rm bulge}/M_{\rm disk}$ criterium:
\textit{yellow} for $M_{\rm bulge}/M_{\rm disk}<0.25$, 
\textit{magenta} for $0.25<M_{\rm bulge}/M_{\rm disk}<0.5$, 
\textit{blue} for $0.5<M_{\rm bulge}/M_{\rm disk}<0.75$ and
\textit{red} for $0.75<M_{\rm bulge}/M_{\rm disk}<1$. The whole sample of spiral galaxies
selected following Hatton et al. (2003) in terms of the relative luminosity of bulge and disk 
is represented by the \textit{cyan} dots. The model 
velocities are taken at the fiducial value of $2.2 \ R_{\rm D}$. 
In all bands, we compare the model predictions with the observational relation by Verheijen (2001),
which is represented by two lines, indicating a different
selection criterium for the observed galaxies. The \textit{green solid line})
represents the complete sample with a measured HI global profile, while the 
\textit{green dashed line}) represent the RC subsample. In the B band, we also plot the 
observational relation from Pierce \& Tully (1992; \textit{thick 
solid black line}). In the I and K bands, we compare the model TF with the
observational relation by Masters et al. (2006, 2008, \textit{black solid line} for the
relation, and \textit{dot-dashed lines} for its scatter). The I-band TF relation by Masters et al. (2006)
is coincident with the relation found by Giovannelli et al. (1997).

For the B band, the plot shows that the model is successful in reproducing the
observed TF slope, expecially for the later-type disk galaxies,  
and lies instead on the bright side in terms of the zero point. The offset with the Pierce \& Tully
relation is substantial, while it amounts to $\sim 0.2-0.4$ mags with respect to the Verheijen relation.
Notice the discrepancy 
between different observational relations as well; in particular, the relation by 
Pierce \& Tully (1992) is on the faint side of the others, and in fact led to a measured
value of the Hubble parameter $H_0 \sim 85$ km/s/Mpc, much larger than more recent results.
For this reason, we consider the discrepancy between the model TF and this particular 
observational relation less significant. 

In the I band, the model is able to reproduce the slope of the I-band TF relation
very well, and is offset in the zero-point by $\sim 0.2-0.4$ mags. The model scatter is
also comparable to the observed one. The performance of the model TF is worse here than the results
of Croton et al. (2006) and de Lucia et al. (2004), who compare their models with
the observed relation of Giovannelli et al. (1997).
Both the Croton et al. (2006) and de Lucia et al. (2004) models differ from ours in the stellar
population models adopted; by using BC03 instead of M05, the TF is shifted towards fainter magnitudes. 
Moreover, their oversimplification of the rotation velocity recipe shifts the TF towards
slower values. 

In the K band, the agreement of the model predictions with the data 
is better than in the other 
bands. Consistently, the K band is less affected by uncertainties such as dust reddening.
Although the model relation is accomodated inside the data scatter, we nonetheless notice a
discrepancy in the slope, which is more evident than in other bands. We can also identify a
cut in velocity, around $log(W) \sim 2.3$, below which the model galaxies are too bright.

\begin{table*}
\caption{$z=0$ model TF relations, power-law fits} 
\label{tableTF0}
\begin{tabular*}{\textwidth}{@{\extracolsep{\fill}}lccccl}
\hline
\noalign{\smallskip}
[$M_{\rm bulge}/M_{\rm disk}$] & [0,0.25] & [0.25,0.5] & [0.5,0.75] & [0.75,1.] & all spirals \\
\noalign{\smallskip}
\hline
\hline
\multicolumn{6}{c}{} \\
B band & & & & & \\
\hline
\noalign{\smallskip}
slope   $\Delta M/\Delta \rm{log}W$ & -7.48 & -7.12 & -6.52 & -5.20 & -4.95 \\
\hline
\noalign{\smallskip}
zero point at $\rm{log}W=2.4$ & -20.76 & -20.48 & -20.15 & -19.91 & -20.18 \\
\hline
\hline
\noalign{\medskip}
I band & & & & & \\
\hline
\noalign{\smallskip}
slope   $\Delta M/\Delta \rm{log}W$ & -7.51 & -6.66 & -6.43 & -5.33 & -5.37 \\
\hline
\noalign{\smallskip}
zero point at $\rm{log}W=2.4$ & -21.50 & -21.33 & -21.12 & -20.94 & -21.11 \\
\hline
\hline
\noalign{\medskip}
K band  & & & & & \\
\hline
slope   $\Delta M/\Delta \rm{log}W$ & -7.97 & -7.02 & -6.74 & -5.72 & -5.71 \\
\hline
\noalign{\smallskip}
zero point at $\rm{log}W=2.4$ & -21.68 & -21.47 & -21.26 & -21.07 & -21.26 \\
\hline
\hline

\end{tabular*}
\end{table*}

Table (\ref{tableTF0}) lists the slopes and zero-points at $\rm{log}W=2.4$ of the model TF relations
shown in Fig.~(\ref{tfz0}), obtained with power-law fits of the kind $M-5 \rm{log} h=a+b \rm{log} W$, 
separated by morphological types. 
Interestingly, in all the bands considered the model shows a net differentiation 
between morphological types, 
with the more disk-dominated galaxies following a steeper relation with brighter zero-points
in all bands, while bulge-dominated objects show a flatter slope.
The morphology dependece of the K-band TF will be investigated in detail later on.

Notice the turn in the model TF slope at the high-mass end, for earlier types; 
this is the combined effect of AGN feedback and bulge formation. 
On the one hand, in the more massive objects the onset of AGN activity 
quenches star formation and thus
reduces the galaxy luminosity per unit mass, moving the objects below the relation.
On the other hand, while at the low-mass end the mechanism of bulge formation is preferentially
secular evolution, which does not introduce a large scatter in the galaxy stellar
populations, at the high-mass end bulges form primarily through mergers, 
which contribute to increase the scatter and raise the mass-to-light ratio. In addition,
the larger bulge masses boost the bulge contribution to the 
rotation curve and increases the rotation velocity. These factors combined push the
model galaxies below the relation at the high-mass end, and increase the scatter in the relation.
This may also be the cause why the velocity range covered by GalICS spirals is not as wide 
as the data. The model in fact does not seem to produce late-type galaxies
that are massive enough (contrary to both the De Lucia et al. 2004 and Croton et al. 2006 models).
If GalICS overpredicts the abundance of large bulges at the high-mass end, then the number
of massive spirals evolving into early-types increases, and these objects are lost from the 
spiral selection. This is mirrored in the progressively
larger scatter and flattening of the slope for earlier-type model spirals.

Overall, although the model produces a reasonable TF slope, 
its performance is unsatisfactory in the B band, it is marginally 
acceptable in the I band, and it is better, but not perfect, in the K band. 
The interplay between dynamics and baryonic physics that shapes the TF relation is 
affected by many factors. We argue that we have good control over the dynamical model
for the galaxy rotation curves and the stellar population models, both of which are physically
sensible. Other degrees of freedom are introduced by the supernovae feedback recipe, 
the chemical evolution model, the star formation history of the model galaxies, dust extiction,
and the modeling of the data to obtain rest-frame quantities (although such adjustments
as the $k$-correction are small for local samples).
The increasingly good match between model and data from the optical to the near-infrared
suggests two possible causes for the offset. 

The first possible explanation is the treatment of dust reddening. The observed relations
used for the comparison have been corrected for internal extinction. Verheijen (2001) argues
that this correction amounts to up to 1.4 mags in the B band. Fluctuations in the dust
corrections can be of the same order of the offset we see between model and data. 
The better agreement between model and data in the K band is therefore particularly significant
in this scenario, since this band is basically unaffected by dust reddening 
(and in addition the $k$-correction 
performed on the data to obtained rest-frame magnitudes is relatively small).
However, if internal extinction corrections were the main cause of the disagreement in the optical TF, 
this would imply that in all the observed relations the dust reddening has been sistematically
underestimated, which seems implausible. 

The disagreement between model and data is present in all bands, but
it is much worse in the optical/blue bands, and presents a morphology
dependence. This suggests that the cause may be found in the balance
between the emissions of stellar populations of different ages, which
means that the star formation histories in the model galaxies are not
realistic. An excess of luminosity like the one we obtain, with a
variance from B to K, may hint to an excessively slow
decline of the star formation rate towards low redshifts. 
As will be shown later on, another hint in favour of this argument is the redshift 
evolution of the TF, the fact that the match between model and data gets better
at higher redshifts.

\subsection{Morphology dependence of the Tully-Fisher at z=0}

From the previous Section it is evident that the ratio $M_{\rm bulge}/M_{\rm disk}$
plays a role in determining the slope of the model TF. In particular, 
as the bulge mass increases with respect to the disk, 
the slope of the model TF decreases and the scatter increases.
In general, model late-types agree better with the observed slope. This is 
encouraging, as the samples of galaxies selected to observe the TF relation 
are usually biased towards late-type spirals, which tend to feature
cleaner rotation curves, due to stronger emission lines (Masters et al., 2008)

\begin{figure*}
\includegraphics[scale=0.7]{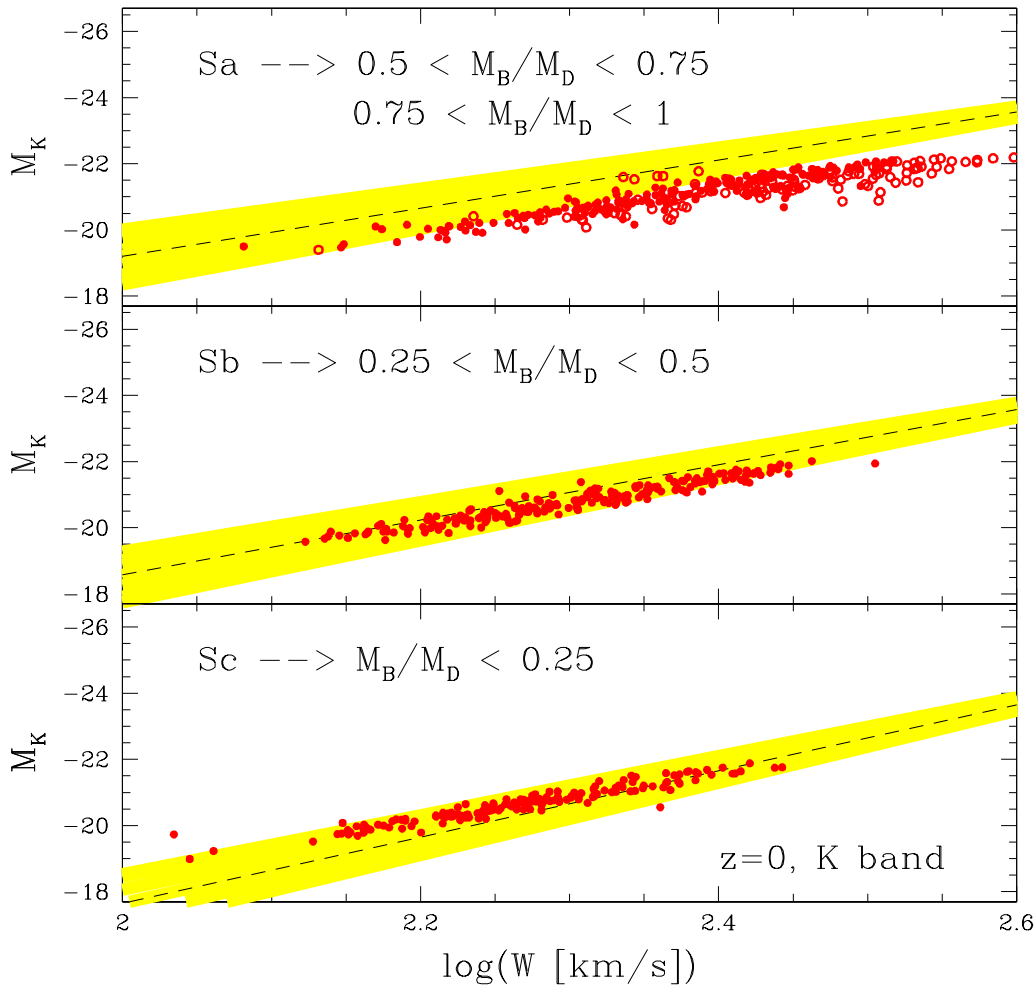}
\caption{The morphology dependence of the $z=0$ K-band Tully-Fisher relation. 
The model predictions are compared with with data from 
Masters et al. (2008) (\textit{dashed solid lines, yellow shaded area representing 
the data scatter}). \textit{From upper to lower}: observed Sa galaxies 
compared with model
galaxies with $0.5 < M_{\rm bulge}/M_{\rm disk} < 0.75$ (\textit{red dots}) and
galaxies with $0.75 < M_{\rm bulge}/M_{\rm disk} < 1$ (\textit{red circles}); 
observed Sb galaxies compared with model galaxies with $0.25 < M_{\rm bulge}/M_{\rm disk} < 0.5$ 
(\textit{red dots}); 
observed Sc galaxies compared with models galaxies with 
$0.25 < M_{\rm bulge}/M_{\rm disk}$ (\textit{red dots}).} 
\label{morph}
\end{figure*}

The $M_{\rm bulge}/M_{\rm disk}$ ratio is a reasonable proxy for Hubble type, in that
it gives a fair representation of the percentage of galaxy mass that is involved
in star formation activity (model bulges do not receive cooling of fresh gas).  
Fig.~(\ref{morph}) shows the morphology dependence of the $z=0$ model 
Tully-Fisher relation in the K band, 
according to our bulge/disk stellar mass classification. The results are shown
in comparison with data from Masters et al. (2008; \textit{dashed black lines and 
yellow shaded area} 
representing the relation and its scatter), and are divided according
to Hubble type. 
In the upper panel, we compare observed Sa galaxies with the model galaxies with  
$0.5<M_{\rm bulge}/M_{\rm disk}<0.75$ (\textit{red dots}), and 
we also show model galaxies with $0.75 < M_{\rm bulge}/M_{\rm disk} < 1$ (\textit{red circles}); 
in the middle panel, we compare Sb types  with 
model galaxies with $0.25<M_{\rm bulge}/M_{\rm disk}<0.5$ (\textit{red dots}); in the lower
panel we compare Sc types with model
galaxies with $0.25<M_{\rm bulge}/M_{\rm disk}$ (\textit{red dots}). 

The model TF shows a differentiation with galaxy morphology, contributed by the
more sophisticated recipe for the model rotation curves. The same differentiation is shown
by the data, with later-type spirals exhibiting a shallower slope. 
The model TF is consistent with the data, with a good match with Masters et al. (2008) 
for Sb/Sc spirals (see Table \ref{tableTF0}), for velocities
$log(W) > 2.2$, although the slope is slightly different and the scatter is smaller.
The model relation for Sa spirals is on the faint side of the data, with a clear bending
of the slope at the high-mass end, 
due to the increasing bulge/disk mass ratio in the model galaxies.
The success of the model in producing a morphology dependence of the TF shows that
the balance between bulge and disk emission in the K band is correctly accounted for
thanks to the use of the M05 stellar populations. Moreover, it also shows that the 
balance between the bulge and disk dynamics is effectively represented by the new
rotation curve recipe.

As evident from Fig.~(\ref{morph}) and the Figures in the previous Section, 
the model TF relation shows a small scatter for late-type galaxies, and an
increasingly large scatter as the $M_{\rm bulge}/M_{\rm disk}$ ratio increases. 
This mirrors the different mechanism for bulge formation in late and early type
spirals. Small bulges are likely to be formed entirely through secular evolution, 
they do not alter the rotation curve in a significant way, and their stellar 
populations are coeval with the ones in the disk. Massive bulges are more likely 
to be formed via mergers, which introduces a large random factor both in the 
ratio $M_{\rm bulge}/M_{\rm disk}$ and in the relative ages and chemical composition 
between disk and bulge stellar populations. The presence of a large
bulge manifests itself with a steepening or a bump in the central part of the 
rotation curve, which increases the rotational velocity at $2.2 \ R_{\rm D}$ at a given 
magnitude. Moreover, the presence of a massive bulge tends to 
dim the galaxy emission per unit mass (with respect to a disk-dominated 
object of the same mass).
In addition to increasingly massive bulges, the downward trend
at the very high-mass end of the theoretical relation, for the earlier types, 
is due to the onset of AGN feedback, that quenches star formation, and makes galaxies less
luminous per unit mass. 

Although we plotted the 
Masters et al. (2008) relation as a straight line, from their paper it is 
evident that this trend is visible in the data as well, and is apparent 
for Sa and Sb types. The model seems to mirror this behaviour, although 
in the data, it is observed at about $log(W) \sim 2.6$, while GalICS produces it
around $log(W) \sim 2.4-2.5$. 
It is evident that GalICS cannot cover the mass range
of the data due in part to the simulation limits; very large spirals are 
extremely rare in nature, and they occur with 0 probability in the simulation. 
Moreover, for increasing masses the model galaxies are more 
likely to experience mergers, which produce larger bulges and 
increase the $M_{\rm bulge}/M_{\rm disk}$ ratio; galaxies therefore are transformed into
earlier types and progressively disappear from the TF relation.

\subsection{Relative contribution of dynamics and stellar population models}

\begin{figure*}
\includegraphics[scale=0.7]{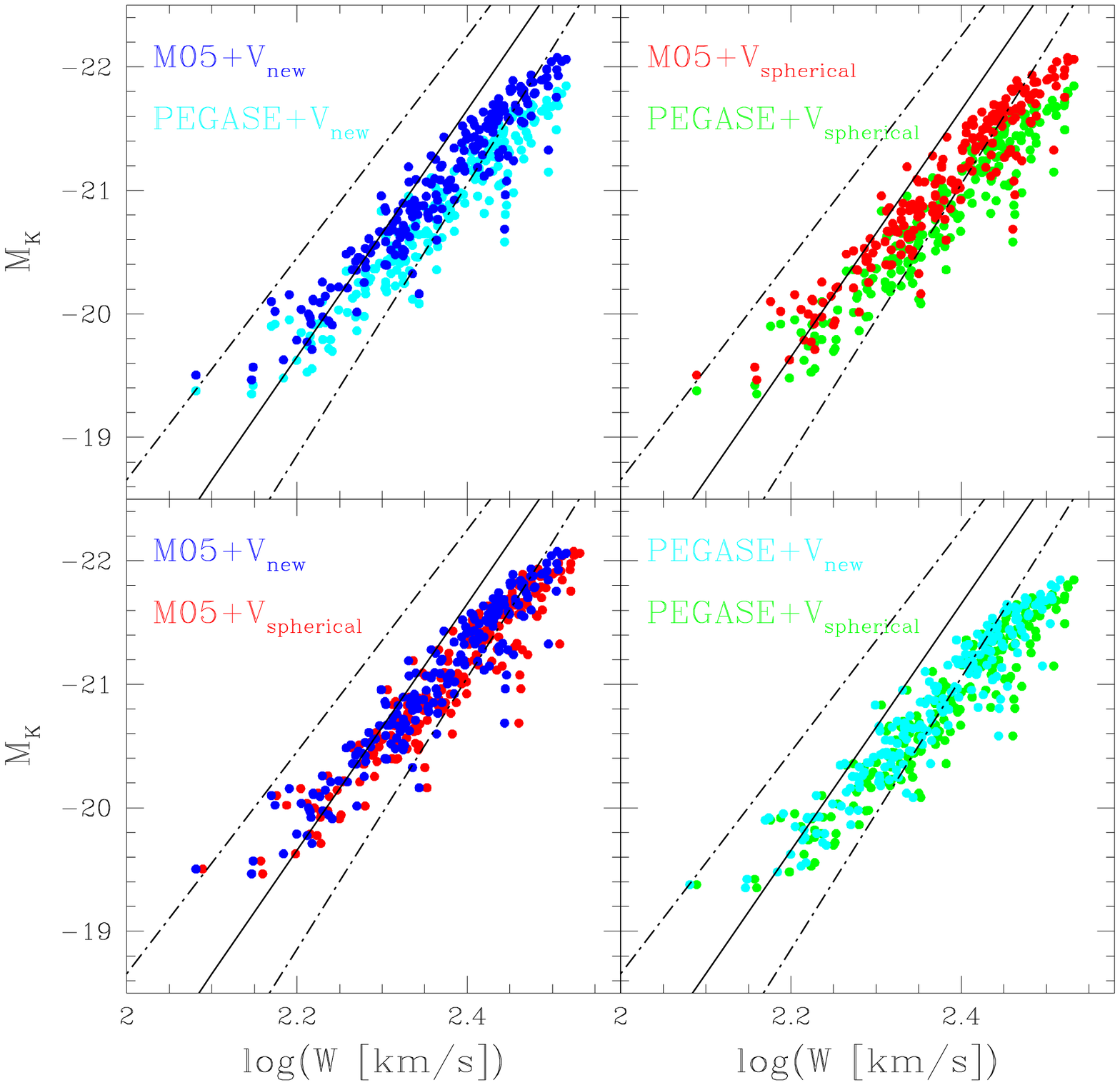}
\caption{Relative contributions of the rotation curve recipe described
in Section 2.1 and the stellar population models described in Section 2.2. 
In the four panels, the $z=0$ model TF relation in the K band is shown for the model
Sa spirals, compared to the corresponding observed TF by Masters et al. (2008) 
(\textit{black solid line}, see lower panel of Fig.~(\ref{morph})). In each panel, 
a different combination of dynamical model $-$ stellar population model is
shown, between the 'new' and 'spherical' model for the rotation curve (see Section 2.1)
and between M05 and PEGASE stellar population models (see Section 2.2), as 
indicated by the labels in each panel.}
\label{contributions}
\end{figure*}

Fig.~(\ref{contributions}) summarizes the relative effects of a more precise and
realistic model rotation curve, and of more complete stellar populations model. 
In the four panels, the $z=0$ model TF relation in the K band is shown for the 
Sa spirals (with errorbars representing its scatter), 
compared to the corresponding observed TF by Masters et al. (2008) 
(\textit{black solid line}, see lower panel of Fig.~(\ref{morph})). In each panel, 
a different combination of dynamical model $-$ stellar population model is
shown, between the 'new' and 'spherical' model for the rotation curve (see Section 2.1)
and between M05 and PEGASE stellar population models (see Section 2.2).
The effect of different rotation curve models and stellar population models on the TF
is of comparable magnitude.
The M05 run of the semi-analytic model produces a K-band TF brighter by $\sim 0.4$ mags
than the PEGASE run. The 'new' recipe for the rotation velocity shifts the TF by 
$\sim 20-30 \%$ in velocity, depending on the galaxy mass and morphology. 
There is a certain degeneracy between the dynamics and stellar
population models. The 'new' dynamical model works better than the 'spherical' 
for both stellar population models. 

The analysis carried out so far implies that the model TF is flawed at $z=0$, 
in the B, I and K bands, even after we adopted refined and updated prescriptions
for the galaxy rotation curves and the stellar populations. 
We use this conclusion to gain a better grasp of more profound problems of the model in reproducing 
the spiral population, in particular regarding the cooling and star formation histories
at low redshifts, as will be discussed further in this work.

\subsection{Satellites and the TF relation}

Satellite galaxies in the simulation are defined as objects that do not reside
in the center of the parent dark matter halo, but are orbiting around or infalling
into a central galaxy. The gas in satellites is stripped because of ram pressure and
tidal effects, while the cooling of new gas is prevented. This effectively shuts down
star formation in satellite galaxies, which evolve passively unless a merger event takes
place. For this reason, model satellite galaxies are ill-suited to be compared
with real spirals. For the purpose of clarification and completeness, we show the 
predicted K-band TF relation for satellite spirals and compare them to centrals.

\begin{figure*}
\includegraphics[scale=0.5]{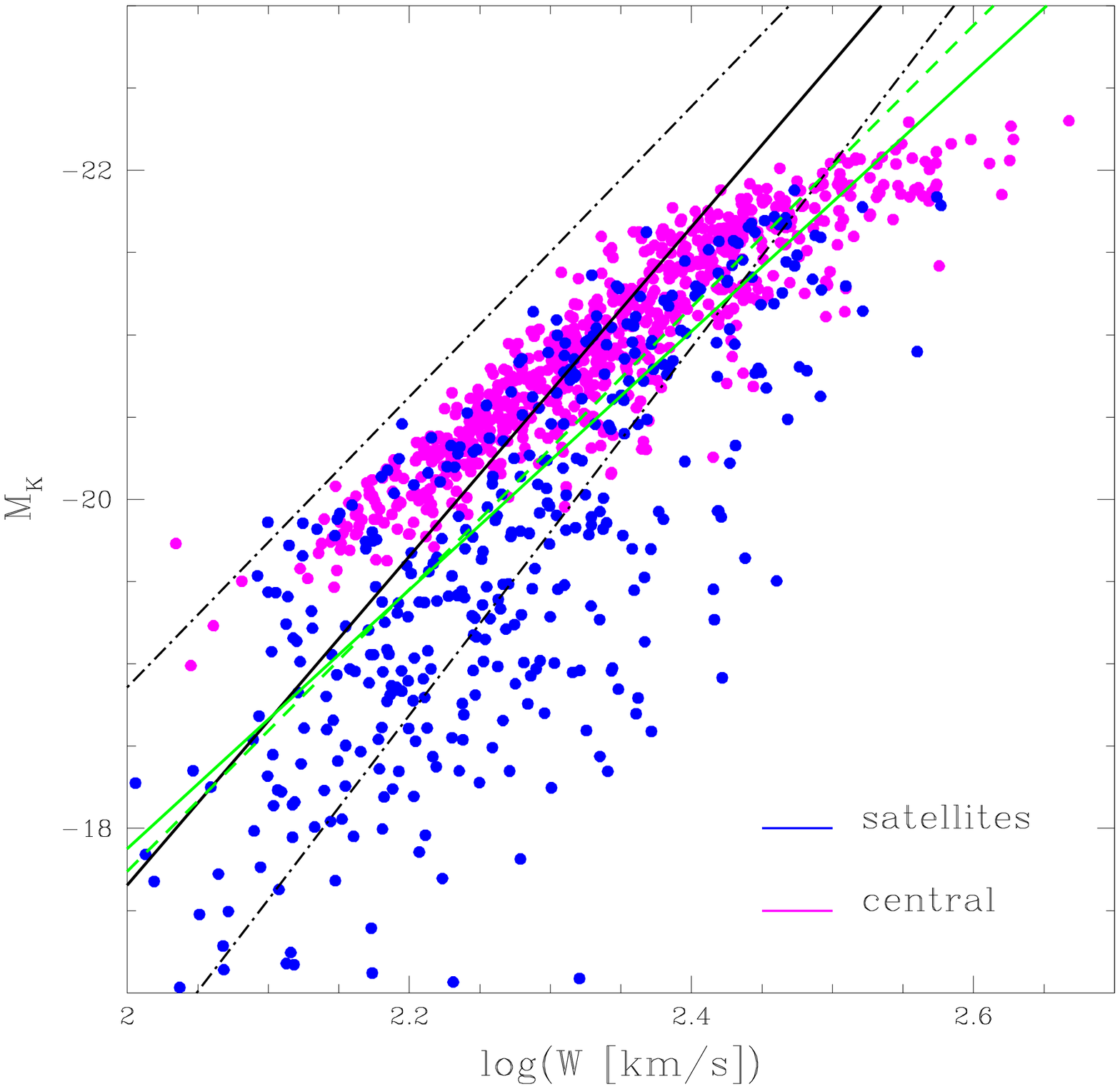}
\caption{The predicted $z=0$ K-band Tully-Fisher relation, for a sample of model spirals
in the range $0 < M_{\rm bulge}/M_{\rm disk} < 1$,
and additionally split into central (\textit{magenta dots}) 
and satellite (\textit{blue dots}) galaxies. 
The model is compared with
the observed relation by Masters et al. (2008; \textit{solid black line} for the
relation, and \textit{dot-dashed lines} for its scatter) and Verheijen 2001
(\textit{green solid/dashed lines}).} 
\label{kz0sat}
\end{figure*}

Fig.~(\ref{kz0sat}) shows the $z=0$ K-band TF relation for a sample of model spiral galaxies 
in the range $0 < M_{\rm bulge}/M_{\rm disk} < 1$, and additionally 
split into central (\textit{magenta dots}) and satellite (\textit{blue dots}) galaxies. 
It is clear that satellites do not follow a TF relation, but scatter below
the central galaxy relation. The main reason is that
satellites are less bright, due to the quenching of the star formation which, even if
not affecting directly the K-band, raises the overall mass-to-light ratio.
This affects expecially the low mass end of the satellite distribution.

Note that a significant fraction of the galaxies in the
Master et al. samples are cluster galaxies, and they
follow the same TF relation as the field samples. 
We can therefore use the TF relation as a tool to evaluate the ability of the
model to reproduce the structure of satellite galaxies. We conclude without doubt that
the model cannot produce realistic satellites. Again, the main problem resides in the 
star formation history. Real satellite spiral galaxies, although perturbed when in dense 
environments, are still star-forming, in particular those selected for TF studies. 
Hierarchical models, in order to reproduce the observed colour-magnitude relation, are
forced to shut down star formation in satellites. This highlights a fundamental problem 
in hierarchical models, that certainly deserves further investigation.

\section{Stellar mass \textit{vs} K-band TF relation: S0 galaxies?}

There is an ongoing debate about the formation of S0 galaxies. The origin of these objects
is still not understood, but current scenarios include massive bulge formation,   
tidal stripping and star formation quenching as possible causes of a transition
from spirals to S0, via a progressive fading of the disk (see Poggianti et al. 2001
and references therein). 
In particular, the possibility that S0 galaxies may
be spiral galaxies where star formation has been shut down, and
the spiral structure has thus become invisible, can be tested with the present study. 
An interesting consequence of this
is the fact that the TF relation should present an offset between S0 and spiral
galaxies (Williams et al. 2009, Bedregal et al. 2006, and references therein), 
which should be visible in the photometric TF but not in the stellar mass-velocity 
TF relation. However,
Williams et al. (2009) find an offset between observed spirals and S0 
both in the K-band TF and in the stellar TF, and therefore argue that S0 galaxies
are not fading spirals. We use our model to verify whether star formation quenching introduces
an offset between the stellar and K-band TF relation.

In this Section we show the model TF relation in the K-band, compared to the 
\textit{stellar} TF relation $M_{\rm star} \ vs \ V_{\rm rot}$, for star-forming and 
nearly-passive disk galaxies at $z=0$. We again differentiate
the galaxy morphology through the $M_{\rm bulge}/M_{\rm disk}$ ratio. Model disk galaxies with
large $M_{\rm bulge}/M_{\rm disk}$ and low star formation rate should in principle
present a spectral energy distribution similar to real S0 galaxies. First, the large 
bulges (built-up preferentially via minor mergers) and the small gaseous mass 
make these objects the transition point between spirals and ellipticals in the model. 
Second, the low star formation rates mimic the quenching due to gas stripping in dense
environments as well as the natural fading due to the exhaustion of the gas reservoir.

\begin{figure*}
\includegraphics[scale=0.8]{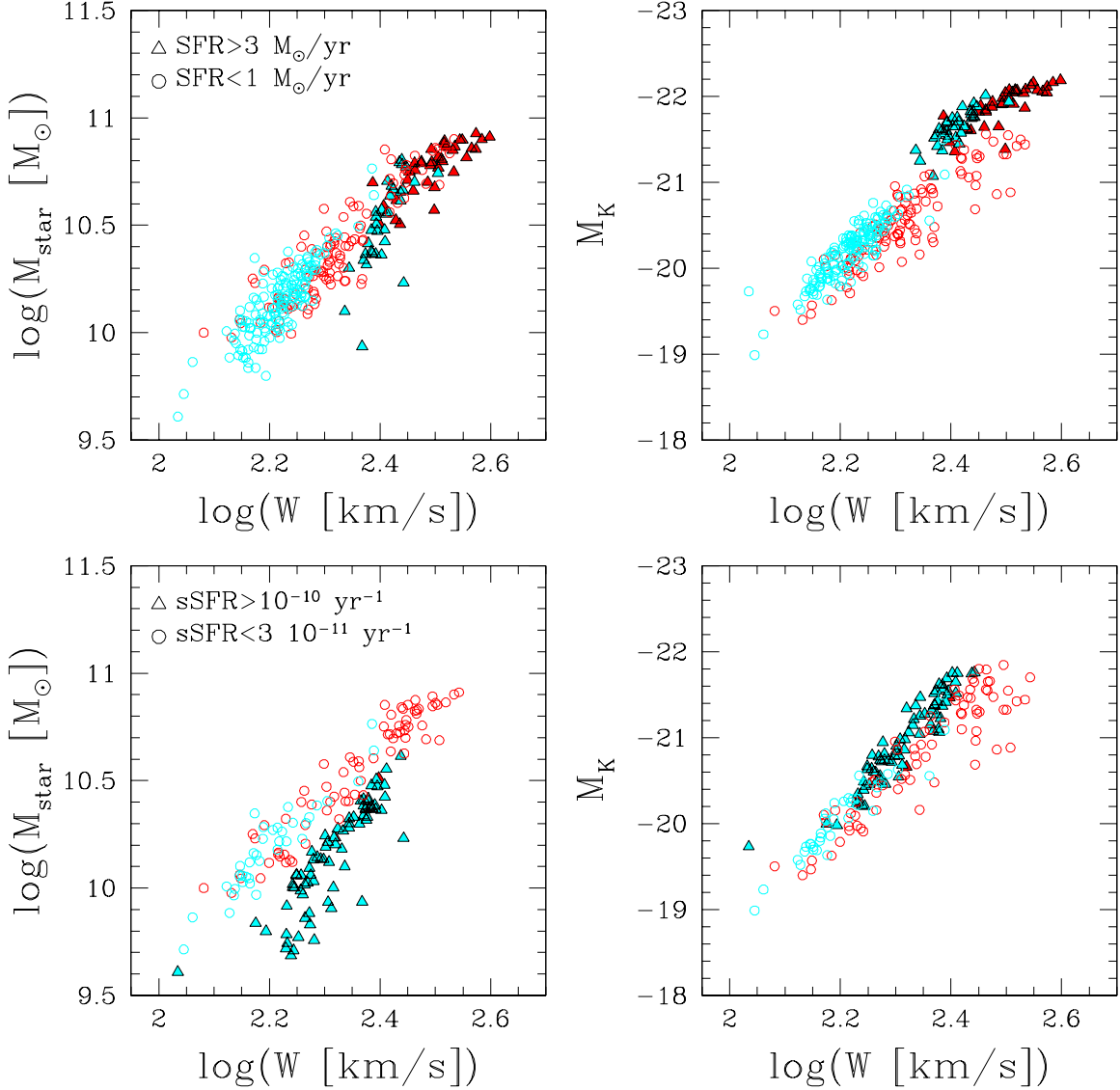}
\caption{Comparison between the stellar and K-band TF relations for star-forming and non-star forming
model galaxies. \textit{Left panels:} model stellar TF relation at $z=0$. \textit{Right panels}: model
K-band TF relation at $z=0$. \textit{Upper panels:} galaxies selected depending on their total
star-formation rate. \textit{Triangles} represent
star forming galaxies with $SFR > 3 \ \rm M_{\odot}/yr$ and \textit{circles} represent nearly-passive
galaxies with $SFR < 1 \ \rm M_{\odot}/yr$.
\textit{Lower panels:} galaxies selected depending on ther specific star-formation
rate. \textit{Triangles} represent
star forming galaxies with $sSFR > 10^{-10} \ \rm yr^{-1}$ and \textit{circles} represent nearly-passive
galaxies with $sSFR < 3 \ 10^{-11} \ \rm yr^{-1}$. 
In all panels, galaxies are colour-coded according to morphology, with
\textit{cyan circles/filled triangles} representing 'later-types' 
with $M_{\rm bulge}/M_{\rm disk}<0.5$, and \textit{red circles/filled triangles} 
representing 'earlier-types' with $0.5<M_{\rm bulge}/M_{\rm disk}<1$.}
\label{tfsfr}
\end{figure*}

Fig.~(\ref{tfsfr}) shows the comparison between the $z=0$ stellar TF (\textit{left panels}) 
and K-band TF (\textit{right panels}) relations, for two subsamples of model galaxies 
divided according to the instantaneous SFR. In the \textit{upper panels}, galaxies are selected 
depending on their total star-formation rate, and 
\textit{triangles} show star-forming
galaxies with $SFR>3 \ \rm M_{\odot}/yr$, while \textit{circles} represent nearly-passive
galaxies with $SFR < 1 \ \rm M_{\odot}/yr$. In the \textit{lower panels} instead, galaxies
are selected according to their specific star formation rates, and \textit{triangles}
represent star-forming galaxies with $sSFR > 10^{-10} \ \rm yr^{-1}$, while 
\textit{circles} represent nearly-passive
galaxies with $sSFR < 3 \ 10^{-11} \ \rm yr^{-1}$ (the separation between star-forming
and nearly-passive was taken from Milky-Way values as a reference,
see Munoz-Mateos et al., 2007). 
In all panels, galaxies are also colour-coded according to morphology, with 
\textit{cyan circles/filled triangles} representing  
'late-type' spirals 
with $M_{\rm bulge}/M_{\rm disk}<0.5$, and \textit{red circles/filled triangles} 
representing 'early-type' spirals 
with $0.5<M_{\rm bulge}/M_{\rm disk}<1$. Red circles therefore represent the model
rendition of S0-like objects (in terms of stellar populations).

When the galaxies are selected according to the total star-formation rate, 
we find a continuous stellar TF relation, as expected, between star-forming and
nearly-passive galaxies, while we find a double sequence in the K-band, with
the star-forming galaxies significantly brighter than their passive counterparts, for
a given mass (the difference in mag is $\sim 0.5$, consistent with the findings of Williams et al.). 
We also find a morphology segregation in the K-band TF, with later-type spirals 
at the brigher end of the population (except the very top of the mass
distribution, where later-types are not found in the model). This segregation tends to disappear
in the stellar TF. This result seems to confirm that, indeed, S0 can be considered quiescent
disks that are in the process of shutting down the star formation, and in this case
the model S0 are represented by the \textit{red circles}. 

If we select the galaxies according to the \textit{specific} star-formation rate,
we find again a different trend between the stellar and K-band TF relations.
The stellar TF relation shows a net segregation between star-forming and nearly-passive
galaxies, which form a separate sequence. In addition, there is a net morphology segregation, 
which puts the earlier-types, nearly-passive galaxies at the massive end of the relation
for any given velocity. In the K-band relation the star-forming, later-type galaxies are clearly
on the brighter, lower-mass end of the relation for any given velocity. If
we consider that the \textit{red circles} as the model S0 galaxies, the picture is consistent
with S0 galaxies being fading spirals. 
Earlier types, characterized by massive bulge formation, occupy the high-mass end of the population. 
The presence of a massive bulge reduces the fraction of galaxy mass that is actively star-forming,
so that the specific star-formation rate and the luminosity are low compared with later-types 
of similar mass. 

Although more investigation is needed on this topic, we are encouraged to conclude 
that the scenario that describes S0 galaxies are fading spirals is promising. 
We note that a direct comparison between our results and the findings of 
Williams et al. (2009) is problematic, due to the differences between the modeling of the dynamical mass 
by Williams et al. and the mass-velocity relation in GalICS. 
Further investigations on these results are under development

\section{The redshift evolution of the TF up to z=1}

The evolution of the TF is a good test of the galaxy assembly mechanism in the 
semi-analytic model, and here we present it in the optical and near-IR.

\subsection{Evolution of the optical TF}
  
\begin{figure*}
\includegraphics[scale=0.9]{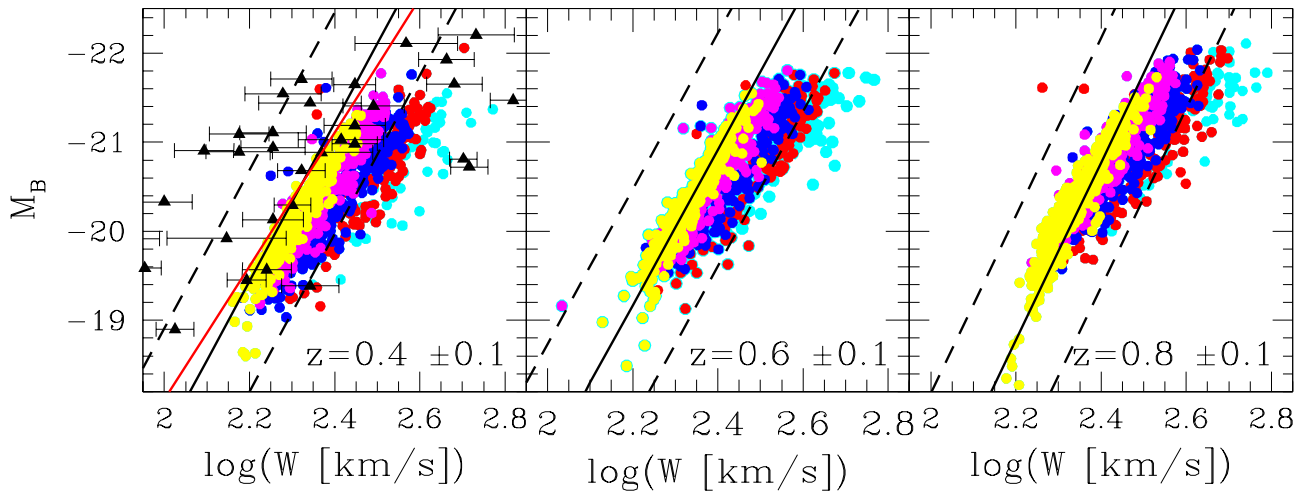}
\caption{The redshift evolution of the rest-frame B-band Tully-Fisher relation from 
$z \sim 0.4$ to $z \sim 0.8$, compared with data from Fernandez-Lorenzo et al. (2009). 
\textit{Solid/dashed black lines} represent the observed relation and its $2 \sigma$ uncertainty, in all
panels. In the $z=0.4$ panel, the data from B\"ohm \& Ziegler (2007) are shown as 
\textit{black triangles}. The relation by Bamford et al. (2006) at $<z> =0.33$ 
is also shown (\textit{red line}). The model galaxies are represented by dots, 
colour-coded as in the previous Figures.}
\label{tfb}
\end{figure*}

Fig.~(\ref{tfb}) shows the redshift evolution of the model rest-frame B-band TF relation from 
$z \sim 0.4$ to $z \sim 0.8$, 
compared with data from Fernandez-Lorenzo et al. (2009; \textit{black solid/dashed lines} representing
the observed relation and its scatter, in all panels), who derive the galaxy rotational velocity
in a somewhat different manner than the rest of the published observational relations, 
based on optical line widths (see also Verheijen 2001 for a comparison between TF relations 
obtained with different velocity estimators). In the $z=0.4$ panel, data from B\"ohm \& Ziegler (2007)
are shown as \textit{black triangles}, and the relation by 
Bamford et al. (2006) at $<z> =0.33$ is also shown for comparison.
The colour-coding of the model dots is as in the previous Figures.

This figure shows a very good agreement between the model late-type spirals 
($M_{\rm bulge}/M_{\rm disk}<0.5$, \textit{yellow and magenta points}) 
and the data at $z \sim 0.6$ and $z \sim 0.8$, in slope,
zero point and scatter. At $z\sim 0.4$ the model reproduces the observed slope, but tends to
be slighly fainter in zero point (although still inside the observed $2 \sigma$ uncertainty).
Again, the model galaxies show a very clear morphological differentiation, and the earlier-types
fall off the relation at the high-mass end at all redshifts, 
which as already discussed, is an effect mainly due to bulge formation, and AGN feedback. 

The decline of the global SFR with time, together with the migration of the
star formation to lower and lower galaxy masses (the downsizing effect), causes the
galaxies to become fainter and the slope of the TF to get shallower at lower redshifts
(also confirmed by the results of Weiner et al. 2006).

The model optical TF agrees very well with the data at high redshift, and by $z \sim 0.4$ the match
starts to fail, where the model galaxies are too faint. By $z \sim 0$, the model is instead
too bright (Fig.~(\ref{tfz0}). This seems to confirm our suspicion that the model galaxy 
star formation histories are the main drivers of the discrepancy (or match) between model and data. 
At $z>0.5$ the model spiral galaxies seem to be quite realistic, while for lower redshifts the 
gas cooling, the accretion of substructures and the supernovae feedback conspire to 
produce an excessive star formation at odds with observations. 

\subsection{Evolution of the near-IR TF relation}

\begin{figure*}
\includegraphics[scale=0.9]{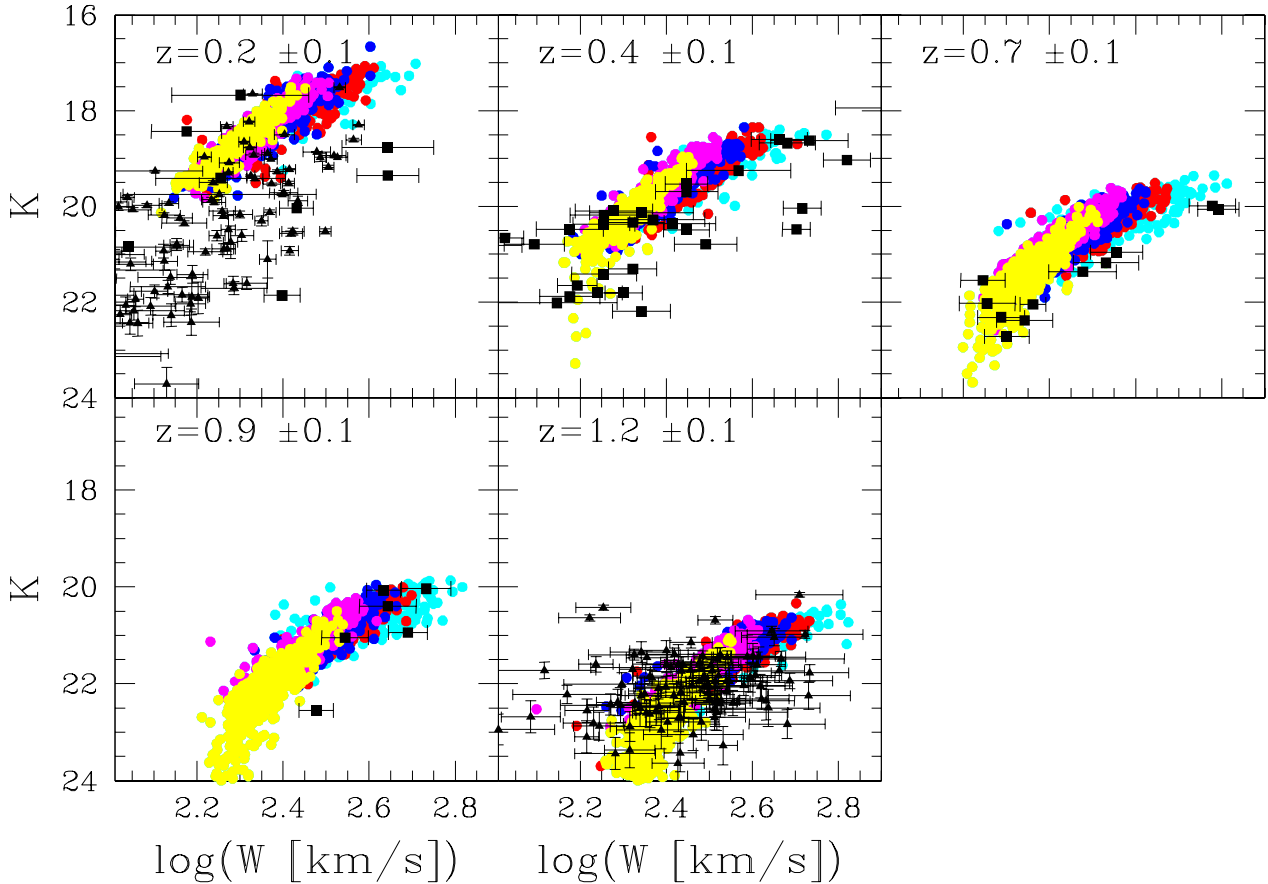}
\caption{The redshift evolution of the observed-frame K-band Tully-Fisher relation from $z \sim 0.2$
to $z \sim 1.2$, 
compared with data from Fernandez-Lorenzo et al. (2010; \textit{solid black triangles}) and 
B\"ohm \& Ziegler (2007; \textit{solid black squares}). The model galaxies are represented by dots, 
colour-coded as in the previous Figures.}
\label{kall}
\end{figure*}

Fig.~(\ref{kall}) shows the evolution of the observed-frame K-band TF relation,
in the redshift range from $z \sim 0.2$ to $z \sim 1.2$. The colour-coding of the model 
galaxies is the same as in the previous Figures. We compare the model predictions with
data from Fernandez-Lorenzo et al. (2010; \textit{solid black triangles}) and B\"ohm \& Ziegler 
(2007; \textit{solid black squares}). In this case the observed magnitudes are not corrected
for dust extinction, so the model galaxy spectra are reddened according
to their star-formation rate, as discussed in Section 2. 

The scatter in the data here is large, and we do not attempt to determine the 
observed slopes and zero points. Nonetheless, the accord between the model TF
and the observations is good in the range $0.4 \le z \le 1.2$. 
Both the observed velocity range and luminosity range are well reproduced.
The model galaxies also occupy the same 
velocity-luminosity space as the data. Moreover, the redshift evolution is  
reasonably well reproduced.

At $z \sim 0.2$, the model galaxies are too bright at the low-mass end, consistently with 
the results at $z=0$ (Fig.~\ref{tfz0}). Following the combined information of the evolution 
of the TF in the optical and near-IR, we argue that   
the model star formation histories are realistic down to $z>0.4$, and odd for lower redshifts.

\section{Discussion}

Disentangling the contributions of the different physical processes conspiring to produce 
the Tully-Fisher is not a trivial task. 
The slope and zero-point of the relation are determined by the
interplay of the dynamics of galaxy assembly inside the dark matter
halo, the cooling that regulates the gaseous and stellar content in
each halo, the mass accretion and star formation histories and the
supernovae feedback, all of which affect the structure and stability
of disks as well as the evolution of the luminosity across all bands.
Thus, matching the observed TF scaling relation is a fundamental check that the model is 
correctly predicting the formation and evolution of spirals.  

However, a meaningful comparison with the observed TF necessitates of some fundamental steps, that 
ensure that there are no systematics that can offset the results. The
first step is the correct modeling of the galaxy rotation curves, which are not
directly predicted by the semi-analytic model itself, but need to be
determined based on the galaxy fundamental dynamical quantities. The
rotation curves extrapolate spatial information in model galaxies,
following a dynamical recipe that is physically motivated and that needs to be as sophisticated as
possible. In our 'new' model, we made use of models for the mass
distribution of the galactic components and angular momentum transfer
to the disk accompanying bulge formation.
It is worth noting that a complementary tool in this sense is
represented by  SPH simulations, which favour a 
detailed study of the galaxy structure, and are able to reproduce the TF relation for single galaxies,
(e.g. Piontek \& Steinmetz, 2009b), even if the shape of the rotation curve often remains unrealistic 
(see for instance Governato et al. 2007), and massive spiral galaxies still represent a problem.
The sophistication of current SPH simulations
also allows for detailed studies of the effects of perturbations such as ram-pressure on the rotation curves 
and the star formation (Kapferer et al. 2009, Kronenberg et al. 2008). 
However, the current resolution limits for this kind of 
simulations does not allow for cosmological runs and a statistical study. 

The second step is the implementation of comprehensive stellar population models, which assures
that the photometric side of the scaling relation is not marred by systematic offsets. We showed that
this element can introduce a bias as large as 1 mag in the K-band, and
0.3-0.4 mag in the I band.

The third step is a consistency check in the comparison with
observations, that presents challenges not related to the
semi-analytic model itself. 
For instance, in a given band, observed TF relations sometimes are not
consistent with one another, and the reason
for these discrepancies is probably to be found in the data analysis. In fact, a significant
degree of modeling of the raw data is required to produce an observed TF. To name a few, 
corrections are required for completeness, internal extinction, inclination, average distance and size 
in case of clusters, morphology, peculiar velocities (see for instance Masters et al. 2006).
In addition, observed-frame magnitudes
have to be converted into rest-frame through $k$-correction.
and this introduces a further bias in the comparison. As we pointed out in Tonini et al. (2010), 
it is preferable to translate the semi-analytic model into observed-frame and compare the 
predictions with the apparent magnitudes. This is crucial expecially for
star-forming objects, because their evolution is fast and the corrections are significant, 
(expecially for internal extinction and $k$-correction).

After the systematics in the modeling and in the comparison with data are taken care of in
'post-production', the hierarchical model can be truly tested, and the
problems of the model in reproducing the TF relation can be
investigated.

\subsection{Star-formation histories}

The offset between the model and the observed TF relation has the
following features: 
\textit{i)} it depends on redshift, with a better agreement between
model and data at higher redshift; \textit{ii)} at redshift $z=0$ it depends on the
photometric band, with a better agreement in the K band and
the worst performance of the model in the B band; \textit{iii)} it
presents a morphology dependence across all redshifts and photometric
bands. 
These facts suggest that the star formation history of the model
galaxies is unrealistic, after $z \sim 0.4$. As a consequence, the 
predicted aging of the stellar populations is off, and the emission of stellar populations of
different ages is unbalanced.
In particular, the decline of star formation with time seems not to be
fast enough, with the consequence that the fraction of the total stellar mass
constituted of young and intermediate-age stellar populations is too
high, a fact that predominantly affects the UV and optical bands, and 
the near-IR via the TP-AGB light. Note that this does not imply a
wrong \textit{instantaneous} SFR at $z=0$, which in fact is predicted
to be consistent with observations (Hatton et al. 2003).

Central galaxies in a hierarchical model
never stop accreting material, including fresh gas and gas previously expelled from the galaxy itself. 
Although this type of star formation history can be chaotic, in most
cases at low redshifts GalICS star formation histories show a
sufficiently steady decline to be approximated in the form of
$\tau-$models: $SFR(t)=SFR_0 exp(-t/t_0)$. We calculated the predicted
luminosity of model galaxies with different star formation timescales,
and found the latter to be a fundamental parameter regulating the relative
luminosities in B and K. For instance,
a galaxy of age of $10 \ \rm Gyr$ with a star-formation history like 
$SFR(t)=SFR_0 exp(-t/10)$ (with a e-folding time $\tau=10 \ \rm Gyr$) produces a 
B-band magnitude brighter by $\sim 0.83$ mags compared to a model like
$SFR(t)=SFR_0 exp(-t/3)$ (with a e-folding time $\tau=3 \ \rm Gyr$), while it
is brighter in the K-band by $\sim 0.46$ mags. From Fig.~\ref{tfz0} it is
evident that a shift of $\sim 0.83$ mags in the B band would produce a much
better agreement between model and data, while at the same time, a shift 
of $\sim 0.46$ mags in the K band would still preserve the accord with the observed
relations.

A faster decline of the star formation in time allows the galaxies to
fade more rapidly at low redshifts, and brings the predicted TF
relation down in luminosity. To size this effect, consider what
happens to satellites after the star formation is shut down
(Fig.~\ref{kz0sat}). Also, as shown in all previous Figures, as more
and more of the fraction of galaxy mass is in the inactive bulge
component, or AGN feedback kicks in, the galaxy tends to fade and fall
below the TF relation. 

We argue that, in order to fix this problem, a more radical revision of gas infall 
and cooling is probably needed. 
A more detailed investigation of this point is of great interest, 
but it is beyond the present scope, in that it requires an extended comparison with SFR derived
from observations at different redshifts. Note that the derivation of SFR from observations
is affected by the adopted stellar population models, dust extinction corrections, 
star-formation history laws, as discussed in Maraston et al. (2010), 
and the use of inhomogeneous sets of such priors in comparison with
our model will lead to significant biases. This line of investigation is currently under 
development

In addition to the star formation history, a second-order effect on the optical TF is 
caused by the chemical evolution model in use. The current version of GalICS implements
the Pipino et al. (2009) recipe which, compared to Hatton et al. (2003), produces slightly
lower total metallicities, causing the galaxies to be brighter. Depending on age and star formation
history, the difference in magnitude amounts to anything between $0$ and $\sim 0.5$ in the B band,
and less for the I and K bands. Accounting 
for uncertainties in the chemical evolution recipe would only lead to a modest increase of  
the TF scatter, which would be consistent with the scatter in the data, but it would not 
shift the overall relation in a significant way.

\subsection{Disk scale-lenghts and the disk-halo connection}

There is also the possibility that part of the discrepancies that we
encountered in our comparison with data are due to a flaw in the
modeling of the galactic dynamics. In particular, the galaxy rotation
may be offset, regardless of the galaxy type, due to a wrong determination
of the galaxy angular momentum. This can be investigate by analyzing
the disk scale-lenghts $R_{\rm D}$ produced by the model.  

\begin{figure}
\includegraphics[scale=0.45]{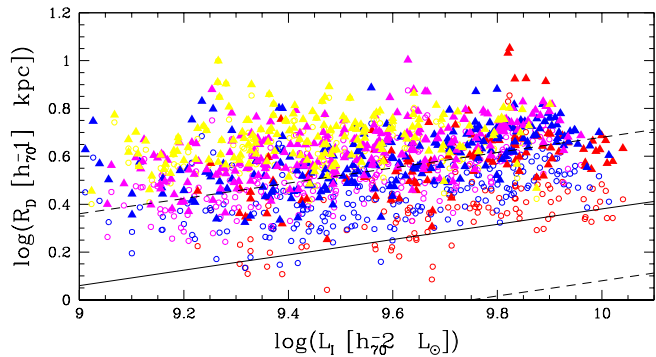}
\caption{Model disk scale-lenghts as a function of I-band luminosity, compared with the 
relation obtained by Courteau et al. (2007) from observations (\textit{solid/dashed lines} for the
relation and its scatter). The \textit{empty dots} represent model galaxies in the 'spheroidal' model,
colour-coded according to $M_{\rm bulge}/M_{\rm disk}$ as in the previous Figures. \textit{Filled triangles}
represent the same galaxies after angular momentum transfer from the bulge is evaluated (the 'new' model).}
\label{RL}
\end{figure}

Figure \ref{RL} shows the model disk scale-lenghts as a function of I-band luminosity, compared with the 
relation obtained by Courteau et al. (2007) from observations (\textit{solid/dashed lines} for the
relation and its scatter). The \textit{empty dots} represent model galaxies in the 'spheroidal' model,
colour-coded according to $M_{\rm bulge}/M_{\rm disk}$ as in the previous Figures. \textit{Filled triangles}
represent the same galaxies after angular momentum transfer from the bulge is evaluated (the 'new' model).

It is evident that the model cannot reproduce the Courteau et al. relation. 
The disk-scale lenght appears to be too large, and the scatter too big, 
the latter a problem that was mentioned already in Hatton et al. (2003). 
More in detail, the model does not reproduce the Courteau et al. relation for the 
late-type spirals (\textit{yellow points}), which are mostly untouched by our 'new' velocity
model. In the old 'spheroidal' model, the earlier-type spirals present
a marginal agreement with the data, which we argue is accidental. In
fact, the values of the disk-scale lenghts inferred from observations
are commonly obtained by fitting the surface brightness profile with an exponential law, and 
that the result is more accurate the closer the galaxy is to a pure disk. Galaxies with large bulges
are not ideal candidates, and they tend to be excluded from the observational samples. 
The 'new' model effectively corrects for the presence 
of the bulge, moving the earlier-type spirals in the same $R_{\rm D}-L_{\rm I}$ space as the later-types, 
in a sense making all galaxies consistently offset with the Courteau et al. relation.

Before analyzing the possible causes of the discrepancy, it should be noted that a source  of error stems from the use of Eq. (\ref{rd}) for
the determination of $R_D$. This formula, adopted by Hatton et
al. 2003, produces the highest $R_D$ under angular momentum
conservation, in the MMW formalism. Additional shape factors that can
reduce the value of $R_D$ are set to unity, but as discussed in Tonini
et al. (2006a), their values as inferred from observed rotation curves
are actually close to unity, and vary slowly with disk mass.

As for the cause of the discrepancy with the Courteau data, we offer 2 explanations: 

I) as discussed in Tonini et al. (2006a) and D'Onghia \& Burkert (2004), 
the halo spin parameter, which is found to vary from $<\lambda>=0.04$ in GalICS to 
$<\lambda>=0.05$ in other simulations, is too high compared with the one
that can be inferred from observations of spirals. 
If $<\lambda>=0.03$ for instance (the value inferred by Tonini et al. 2006a), 
then Rd is smaller by $ \sim 25\%$, which would push our model $R_{\rm D}$ towards the Courteau et al.
relation. Note that, although $\lambda$ originates naturally from tidal forces during structure formation, 
the discrepancy between the simulated and observed $\lambda$ is not necessarily due to a flaw in the
simulations, but rather it hints at a subsequent redistribution of angular momentum between baryons and
dark matter;
 
II) as mentioned in Section (2.1), the ratio
$j_{\rm D}/m_{\rm D}$ is equal to unity only under angular momentum conservation, 
which is not a realistic
scenario during galaxy formation. In particular, if the baryonic
component that collapses to form the protogalaxy is clumpy, it likely
dissipates angular momentum in its path to the centre of the dark
matter halo, as proposed by Tonini et al. (2006b), and therefore $j_{\rm D}/m_{\rm D}<1$. 
Consequently, the disk scale-length is smaller, and
the tension between model $R_D$ and observations is partially
alleviated ($j_{\rm D}/m_{\rm D} \sim 0.75$
leads to a decrease of $R_{\rm D}$ by $\sim 25\%$). The MMW recipe can therefore be considered a 'maximal scale-length scenario', with the real scale-length distribution being centered on smaller values of $R_D$ because of angular momentum transfer. 

Both the effects described above contribute in reducing the size of
$R_D$, and are currently not investigated in semi-analytic
models. Note also that such effects do not depend on redshift nor on
photometric band, therefore they do not affect our conclusions on the
model star formation histories discussed in the previous section.

An additional, largely unknown source of scatter, on both the disk
scale-lengths and the rotation curve, is represented by
the dark matter halo structural evolution. On the one side,
readjustments in the halo profile induced by the baryon infall can
lead to a dramatic change in the inner density profile (Tonini et
al. 2006b), thus altering the contribution of the halo to the total
velocity. On the other side, tidal interactions between halos and
mass accretion in the hierarchical assembly repeatedly strip and add
material in each halo. There is no standard recipe in semi-analytic
models to recalculate the halo equilibrium structure that takes into
account all these effects, and since there is no spatial information
inside single objects, the mass is the primary driver of halo
evolution. The virial radius in particular is not recalculated after
every interaction, with the consequence that the halo density profile
can oscillate quite a lot. Note that also in N-body simulations,
although all halos are fitted to a common functional profile (like the
NFW or the isothermal sphere), the
scatter on the profile is very large, as is the scatter on the
mass-virial radius relation (Bullock et al. 2001). The scatter in the virial radius 
directly translates into a scatter in $R_D$, while the scatter in
the density profile affects the halo velocity. In particular, an overestimation of
the disk scale-lengths will occurr in cases where the mass loss via
tidal stripping dominates the halo evolution. 

The offset between model and observed disk scale-lengths represents an
interesting failure of the model, that arguably sheds light on some missing physics.
These considerations suggest that baryon cooling and halo structural
evolution in the model need some more fundamental revision, 
which is beyond the scope of the present work. 

\section{Summary and conclusions}

We analyzed the predictions of the hierarchical galaxy formation model GalICS 
on the Tully-Fisher relation and its evolution with redshift. 
We introduced two new elements in the model: 1) a new recipe for the determination
of the galaxy rotation curves, based on the correct dynamical modeling of the different
galactic components, and taking into account the angular momentum exchange between bulge
and disk due to secular evolution; 2) the M05 stellar population models, that 
include an exhaustive treatment of the TP-AGB emission, which affects galaxies with ongoing star formation. 

Our main conclusions are: 

$\bullet$ the improvement on the dynamical description of disk galaxies and on the stellar
population models impacts the model TF by effects of comparable magnitude; in particular, 
velocities are shifted to values lower by $\sim 0-50 \%$ depending on the galaxy mass and
morphology, and the K-band magnitudes are brighter by $\sim 0.2$ mags at redshift zero and
up to $> 1$ mags at redshift $z \sim 1$ and above;

$\bullet$ at redshift $z=0$ the model reproduces the slope of the observed TF for Sb/Sc 
spirals, in the B, I and K band, but not the zero-point, which is too
bright in all bands. In particular, in the K band the zero-point is
too bright for Sb/Sc galaxies at the low-mass end ($log(W)<2.3$),
although the model galaxies lie within the scatter of the observed TF relation.
In the B and I band, the predicted zero-point for Sb/Sc galaxies is
too bright across the mass range. We argue that the most probable
cause of the discrepancy lies in unrealistic star-formation histories
at $z<0.4$; 

$\bullet$ the model predicts a morphology dependence of the TF relation at all redshifts. 
At $z=0$ observations in the K-band confirm this trend, and the comparison between model
and data is encouragingly good, expecially for Sb/Sc spirals. 
This is an important confirmation that our more sophisticated treatment of the galaxy dynamics
is realistic. The model rotation curve now mirrors the morphological
differentiation observed in nature, well representing the balance
between bulge and disk, and the consequent variations in the
predicted TF along the Hubble sequence are put in sharper evidence. 
The model predicts a steeper slope and a smaller scatter for
later-type galaxies, and a progressive flattening of the slope and
larger scatter for earlier types. This is coherent with the picture of bulge formation in
a hierarchical scenario, where the incidence of mergers is significant in the case of massive bulges, 
while small bulges are more likely to develop via secular evolution. 
Moreover, with the M05 stellar population models, 
GalICS is able to correctly reproduce the balance between the bulge and disk emission in the K band;

$\bullet$ the scatter in the model galaxies increases from late types (Sb/Sc) to early types (Sa/S0),
because the main mechanism for bulge formation switches from secular evolution (Sb/Sc) to minor
mergers (Sa/S0). At the high-mass end, earlier-type galaxies show a flattening of the slope and
tend to fall below the relation due to the presence of massive bulges
and AGN feedback;

$\bullet$ the study of the TF relation for model satellites galaxies
confirms previous hints that the model cannot produce realistic
non-central spirals;

$\bullet$ the TF relation can be used as a discriminating tool to
investigate the origin of S0 galaxies, in the scenario where these
objects form from the fading of spirals after star-formation shutdown;  

$\bullet$ the model reproduces the redshift evolution of the TF for later-type galaxies 
from $z=0.8$ to $z=0.4$ in the rest-frame B band, where slope, zero-point and scatter are
reproduced for Sb/Sc spirals. The model is consistent with the data in the observed-frame 
K band at redshifts between $z=1.2$ and $z=0.4$. 
The simultaneous match in the optical and near-IR evolution, although needing more tests
with larger data samples, indicates that the mass assembly and star
formation histories, as well as the supernovae feedback
implementation, are satisfactory at these redshifts;

$\bullet$ the model cannot reproduce the whole optical Tully-Fisher evolution
from $z \sim 0$ to $z \sim 1.2$. We argue that the decline of the star
formation is too slow, thus preventing the model from matching
simultaneously the local and distant relations;

$\bullet$ the model produces disk scale-lengths too large compared
with observations. We argue that this mirrors a too simplistic recipe
of disk formation inside dark matter halos, that does not take into
account angular momentum redistribution during the baryonic collapse.

\section*{Acknowledgments}
The authors wish to thank the Referee for her/his very useful comments and suggestions, which 
contributed to improve this work. 
CT and CM acknowledge the Marie Curie Excellence Team Grant ''Unimass'' (MEXT-CT-2006-042754)
of the Training and Mobility of Researchers programme financed by the European Community. 
The authors wish to thank Bruno Henriques, Karen Masters,
Michele Cappellari, Michael Bureau and Susan Kassin for their interesting suggestions
and comments.

\end{document}